\documentclass[final,5p,times,twocolumn]{elsarticle}

\usepackage[T1]{fontenc}
\usepackage[utf8]{inputenc}

\usepackage{amsmath}
\usepackage{amsfonts}
\usepackage{amssymb}
\usepackage{amsxtra}
\usepackage{array}
\usepackage{color}
\usepackage{dcolumn}
\usepackage{enumitem}
\usepackage{graphicx}
\usepackage{hepunits}
\usepackage{xspace}
\usepackage{CJKutf8}

\definecolor{purple}{rgb}{0.5,0,0.5}
\definecolor{blue}{rgb}{0.0,0,0.9}
\definecolor{prdblue}{rgb}{0.133,0.118,0.498}
\usepackage[colorlinks=true, pdfstartview=FitV, linkcolor=prdblue, citecolor= prdblue, urlcolor=prdblue]{hyperref}

\usepackage[mathscr,scaled=1.15]{urwchancal}
\DeclareFontFamily{OT1}{pzc}{}
\DeclareFontShape{OT1}{pzc}{m}{it}%
{<-> s * [1.15] pzcmi7t}{}
\DeclareMathAlphabet{\mathpzc}{OT1}{pzc}{m}{it}

\biboptions{sort&compress}

\journal{Physics Letters B}

\begin{document}
\begin{CJK}{UTF8}{song}

\begin{frontmatter}

%\title{Projections for pion and kaon GPDs}
\title{$\,$\\[-7ex]\hspace*{\fill}{\normalsize{\sf\emph{Preprint no}. NJU-INP 073/23, USTC-ICTS/PCFT-23-13}}\\[1ex]
Polarised parton distribution functions and proton spin}

\author[NJU,INP]{P. Cheng% (程鹏)%
        $\,^{\href{https://orcid.org/0000-0002-6410-9465}{\textcolor[rgb]{0.00,1.00,0.00}{\sf ID}}}$}

\author[NJU,INP]{Y. Yu% (俞杨)%
        $\,^{\href{https://orcid.org/0009-0008-8011-3430}{\textcolor[rgb]{0.00,1.00,0.00}{\sf ID}}}$}

\author[NJU,INP]{H.-Y. Xing% (邢惠瑜)%
    $\,^{\href{https://orcid.org/0000-0002-0719-7526}{\textcolor[rgb]{0.00,1.00,0.00}{\sf ID}}}$}

\author[USTC1,USTC2]{C. Chen% (陈晨)%
        $\,^{\href{https://orcid.org/0000-0003-3619-0670}{\textcolor[rgb]{0.00,1.00,0.00}{\sf ID}}}$}

\author[NJU,INP]{Z.-F. Cui% (崔著钫)%
    $\,^ {\href{https://orcid.org/0000-0003-3890-0242}{\textcolor[rgb]{0.00,1.00,0.00}{\sf ID}}}$}
%\email[]{cdroberts@nju.edu.cn}
%\ead{phycui@nju.edu.cn}

\author[NJU,INP]{C. D. Roberts%
       $^{\href{https://orcid.org/0000-0002-2937-1361}{\textcolor[rgb]{0.00,1.00,0.00}{\sf ID}},}$}
%\author[NJU,INP]{Craig D. Roberts}
%\email[]{cdroberts@nju.edu.cn}
%\ead{cdroberts@nju.edu.cn}

\address[NJU]{
School of Physics, Nanjing University, Nanjing, Jiangsu 210093, China}
\address[INP]{
Institute for Nonperturbative Physics, Nanjing University, Nanjing, Jiangsu 210093, China}
\address[USTC1]{Interdisciplinary Center for Theoretical Study, University of Science and Technology of China, Hefei, Anhui 230026, China}
\address[USTC2]{Peng Huanwu Center for Fundamental Theory, Hefei, Anhui 230026, China
\\[1ex]
%
%\hspace*{-8em}Email addresses:
%\hspace*{8em}
\href{mailto:chenchen1031@ustc.edu.cn}{chenchen1031@ustc.edu.cn} (C. Chen);
\href{mailto:phycui@nju.edu.cn}{phycui@nju.edu.cn} (Z.-F. Cui);
\href{mailto:cdroberts@nju.edu.cn}{cdroberts@nju.edu.cn} (C. D. Roberts)
\\[1ex]
Date: 2023 Apr 24\\[-6ex]
%Date: 2023 Apr 10\\[-6ex]
}

\begin{abstract}
Supposing there exists an effective charge which defines an evolution scheme for both unpolarised and polarised parton distribution functions (DFs) that is all-orders exact and using \emph{Ans\"atze} for hadron-scale proton polarised valence quark DFs, constrained by flavour-separated axial charges and insights from perturbative quantum chromodynamics, predictions are delivered for all proton polarised DFs at the scale $\zeta_{\rm C}^2 = 3\,$GeV$^2$.  The pointwise behaviour of the predicted DFs and, consequently, their moments, compare favourably with results inferred from data.  Notably, flavour-separated singlet polarised DFs are small.  On the other hand, the polarised gluon DF, $\Delta G(x;\zeta_{\rm C})$, is large and positive.  Using our result, we predict $\int_{0.05}^1\,dx\,\Delta G(x;\zeta_{\rm C}) = 0.214(4)$ and that experimental measurements of the proton flavour-singlet axial charge should return $a_0^{\rm E}(\zeta_{\rm C}) = 0.35(2)$.
\end{abstract}


\end{frontmatter}
\end{CJK}

\section{Introduction}
%%\alinea
%\label{sec:Intro}
%\noindent\textbf{1.$\;$Introduction}.
%
The proton is Nature's most fundamental bound-state.  In isolation, it is stable; at least, the lower bound on its lifetime is many orders-of-magnitude greater than the $\sim 14$-billion-year age of the Universe.  Further, the proton is characterised by two basic Poincar\'e invariant quantities: mass squared, $m_p^2$; and total angular momentum squared, $J^2=J(J+1)=3/4$.  One can also include parity, $P=+1$, in which case the proton is identified as a $J^P=\tfrac{1}{2}^+$ state.

Contemporary theory posits that the proton is constituted from three valence quarks: $u+u+d$, which interact according to rules laid down by the Lagrangian density of quantum chromodynamics (QCD).  Itself, QCD is a Poincar\'e-invariant quantum non-Abelian gauge field theory.  It is worth stressing that $P$ is a Poincar\'e invariant quantum number.  On the other hand, every separation of $J$ into a sum of orbital angular momentum and spin, $L+S$, is observer dependent.  Hence, there is no relation between $P$ and $L$ in QCD and no objective (Poincar\'e-invariant) meaning for $L$, $S$ separately \cite{Brodsky:2022fqy}.

Assuming isospin symmetry, \emph{viz}.\ that $u$ and $d$ quarks are mass-degenerate, then the wave function of the $J^P=\tfrac{1}{2}^+$ proton is a Poincar\'e-covariant four-component spinor whose complete form involves 128 distinct scalar (Poincar\'e-invariant) functions \cite{Eichmann:2009qa, Wang:2018kto}.  It follows that in any observer-dependent reference frame, this wave function contains $\mathsf S$-, $\mathsf P$- and $\mathsf D$-wave orbital angular momentum components.  The character of the angular momentum described by these components depends on the degrees-of-freedom (dof) used to solve the proton bound state problem.  Typically, those dof change with the resolving scale of any probe used to measure a proton property.
Evidently in QCD, there is no scale at which the proton $J=\tfrac{1}{2}$ can simply be the sum of the spins of the valence dof \cite{EuropeanMuon:1987isl}.
Analogous statements may be made about $m_p^2$; namely, the manner by which the proton mass is shared amongst its constituents depends upon the choices of variables and frame made when solving the bound-state problem.  Either or both of these may depend on the resolving scale used to specify the problem.

These remarks make plain that there is no objective meaning to any separation of the proton's $J=\tfrac{1}{2}$ into subcomponents of any kind.  Such a separation is contextual, acquiring significance only once choices of variables and frame are made.  A useful frame is that reached after projection of Poincar\'e-covariant wave functions onto the light-front because the wave functions obtained thereby are the probability amplitudes connected with parton distribution functions (DFs) \cite{Brodsky:1989pv, Brodsky:1997de, Heinzl:2000ht}.

Issues related to the choice of variables are more complex.  Herein we adopt a perspective characteristic of continuum Schwinger function methods (CSMs) \cite{Eichmann:2016yit, Qin:2020rad}.  Namely, at the hadron scale, $\zeta_{\cal H}<m_p$, QCD bound-state problems are most efficiently solved in terms of dressed-parton dof: dressed-gluons and -quarks, each of which possesses a momentum-dependent mass.  This approach is firmly founded in QCD theory and has widely been used with phenomenological success -- see, \emph{e.g}., Refs.\,\cite{Roberts:2021nhw, Binosi:2022djx, Roberts:2022rxm, Papavassiliou:2022wrb, Ding:2022ows, Salme:2022eoy, Ferreira:2023fva, Carman:2023zke} for discussions of both facets.  Notably, $\zeta_{\cal H}$ is the scale at which all properties of a given hadron are carried by its valence quasiparticle dof \cite{Ding:2019qlr, Cui:2020dlm, Cui:2020tdf}.

\section{Proton Faddeev equation}
\label{PFE}
A key feature of strong interactions is emergent hadron mass (EHM), \emph{i.e}., absent Higgs boson couplings into QCD, the dynamical generation of a nuclear-scale mass for the proton, $m_p\approx 1\,$GeV, concomitant with the formation of massless pseudoscalar Nambu-Goldstone bosons \cite{Roberts:2016vyn}.  As a corollary of EHM, any quark+antiquark interaction that delivers a good description of meson properties also generates nonpointlike quark+quark (diquark) correlations in multiquark systems \cite{Cahill:1987qr}.  A discussion of the empirical evidence for such diquark correlations is presented elsewhere \cite{Barabanov:2020jvn}.   Profiting from the emergence of diquarks, a fully-interacting dressed-quark+nonpointlike-diquark approximation to the proton bound-state equation was introduced in Refs.\,\cite{Cahill:1988dx, Reinhardt:1989rw, Efimov:1990uz}.  It has proven effective in describing proton properties -- see, \emph{e.g}., Ref.\,\cite[Sec.\,2.2]{Barabanov:2020jvn}.  A benefit of working with this simplification is that at most 16 (instead of 128) scalar functions are necessary to fully express the Poincar\'e-covariant proton wave function.

\begin{table}[t]
\caption{\label{anValence}
Gegenbauer coefficients that define the hadron-scale valence quark DFs in Eq.\,\eqref{ExpandGegenbauer}.
Each entry should be divided by $10^3$.
%Each $n\geq 3$ entry should be divided by $10^4$.
 }
\begin{center}
\begin{tabular*}%{|c|c|c|c|c|c|c|}\hline
{\hsize}
{
c@{\extracolsep{0ptplus1fil}}
c@{\extracolsep{0ptplus1fil}}
c@{\extracolsep{0ptplus1fil}}
c@{\extracolsep{0ptplus1fil}}
c@{\extracolsep{0ptplus1fil}}}\hline\hline
\multicolumn{5}{c}{$a_n^{\mathpzc u}$}\\\hline
1 & 2 & 3 & 4 & 5 \\
$403$ & $112$ & $7.31 $ & $-10.4$ & $-4.90\ $ \\\hline
6 & 7 & 8 & 9 & 10 \\
$-0.0474$ & $1.15$ & $0.828$ & $0.334$  & $0.0635$ \\\hline\hline
\multicolumn{5}{c}{$a_n^{\mathpzc d}$}\\\hline
1 & 2 & 3 & 4 & 5 \\
$482$ & $161$ & $20.8$ & $-11.5$ & $-7.11\ $ \\\hline
6 & 7 & 8 & 9 & 10 \\
$-0.430$ & $1.50$ & $1.03$ & $0.349$  & $0.0543$ \\\hline\hline
\end{tabular*}
\end{center}
\end{table}
%{1., 0.402841, 0.111809, 0.00730752, -0.0104338, -0.00489712, \
% -0.0000473794, 0.00114772, 0.00082797, 0.000334312, 0.0000635114}
% {1., 0.482089, 0.160647, 0.0208211, -0.0115301, -0.00710802, \
% -0.000430068, 0.00149911, 0.00102581, 0.000349144, 0.0000543158}

In proceeding, the following hadron-scale quark+diquark Faddeev equation predictions are important.
\begin{enumerate}[label=(\emph{\roman*}), ref=(\roman*)]
\item The proton contains both isoscalar-scalar (SC) and isovector-axialvector (AV) diquark correlations, with the AV correlations being responsible for $\approx 35$\% of the wave function canonical normalisation \cite{Mezrag:2017znp}.  \label{item2i}
\item In the rest frame, the proton quark+diquark Faddeev wave function contains \mbox{$\mathsf S$-}, $\mathsf P$- and $\mathsf D$-wave orbital angular momentum components \cite[Fig.\,3a]{Liu:2022ndb}.
\item Calculated unpolarised valence quark DFs in a proton with the preceding two features are reliably interpolated by a finite sum of Gegenbauer polynomials \cite{Chang:2022jri}:
\begin{equation}
\label{ExpandGegenbauer}
{\mathpzc q}(x;\zeta_{\cal H}) = n_{\mathpzc q} 140 x^3(1-x)^3 \left[1 + \sum_{n=1}^{10} a_n^{\mathpzc q} C_n^{7/2}(1-2x)\right]\,,
\end{equation}
$n_{\mathpzc u}=2n_{\mathpzc d}=2$, where the coefficients are listed in Table~\ref{anValence}.
\item In agreement with experiment \cite{Workman:2022ynf}, the predicted proton axial charge is $g_A=1.25(3)$ \cite{ChenChen:2022qpy}, where the uncertainty reflects that on the masses of the SC and AV diquarks: $m_{\rm SC}\approx 0.8\,$GeV; $m_{\rm AV} \approx 0.9\,$GeV.  Furthermore, the $u$ quark fraction of the axial charge is $g_A^u/g_A=0.76(1)$; that of the $d$ quark is $g_A^d/g_A=-0.24(1)$; and $g_A^d/g_A^u=-0.32(2)$. \label{gAvalues}
\item The singlet axial charge of the proton is \cite[Sec.\,9]{Ding:2022ows}:
\begin{equation}
\label{isoscalaraxialcharge}
a_0 = 0.65(2)\,.
\end{equation}
At $\zeta_{\cal H}$, the remainder of the proton spin is stored in quark+diquark orbital angular momentum.  Similar statements hold even if the quark+quark interaction is momentum-independent \cite{Cheng:2022jxe}.  \label{itemv}
\end{enumerate}
%
%(\emph{i}) The proton contains both isoscalar-scalar (SC) and isovector-axialvector (AV) diquark correlations, with the AV correlations being responsible for $\approx 65$\% of the wave function canonical normalisation \cite{Mezrag:2017znp}.
%
%(\emph{ii}) In the rest frame, the proton quark+diquark Faddeev wave function contains \mbox{$\mathsf S$-}, $\mathsf P$- and $\mathsf D$-wave orbital angular momentum components \cite[Fig.\,3a]{Liu:2022ndb}.
%
%(\emph{iii}) In agreement with experiment, the proton axial charge is $g_A=1.25(3)$ \cite{ChenChen:2022qpy}, where the uncertainty reflects that on the masses of the SC and AV diquarks.  Furthermore, the $u$ quark fraction of the axial charge is $g_A^u/g_A=0.76(1)$; that of the $d$ quark is $g_A^d/g_A=-0.24(1)$; and $g_A^d/g_A^u=-0.32(2)$.
%
%(\emph{iv}) The singlet axial charge of the proton is \cite[Sec.\,9]{Ding:2022ows}:
%\begin{equation}
%a_0 = 0.65(2)\,.
%\end{equation}

\section{Polarised valence quark distribution functions at $\zeta_{\cal H}$}
To deliver a Faddeev equation based prediction for the hadron-scale polarised valence quark DFs, one must generalise the methods used for the unpolarised DFs in Ref.\,\cite{Chang:2022jri}, centred on the vector current, to the axial current case.  A symmetry-preserving axial current appropriate for use with a solution of the quark+diquark Faddeev equation has recently been derived \cite{Chen:2020wuq, Chen:2021guo}; but some time will be required before it can be adapted to the calculation of polarised valence quark DFs.

\begin{table}[t]
\caption{\label{gammavalues}
Referring to Eq.\,\eqref{ratiofunctions}, central values of the mapping parameters, $\gamma^{\mathpzc q}$, that reproduce the given quark's contribution to the proton axial charge.
 }
\begin{center}
\begin{tabular*}%{|c|c|c|c|c|c|c|}\hline
{\hsize}
{
l@{\extracolsep{0ptplus1fil}}
c@{\extracolsep{0ptplus1fil}}
c@{\extracolsep{0ptplus1fil}}
c@{\extracolsep{0ptplus1fil}}
c@{\extracolsep{0ptplus1fil}}}\hline
$i$ & 1 & 2 & 3 & 4 \\\hline
$\gamma_i^{\mathpzc u}$ & $0.350$ & $0.621$ & $ 0.575$ & $0.853\ $\\
$\gamma_i^{\mathpzc d}$ & $1.47\phantom{0} $ & $1.26\phantom{1} $ & $2.62\phantom{5} $ & $1.60\phantom{3}\ $ \\\hline
\end{tabular*}
\end{center}
\end{table}
% {fratioO[x, 0.35], fratio[x, 0.6214], fratioN[x, 0.5752], fratioP[x, 0.8526]}
% Taa = {-fratioO[x, 1.474], -fratio[x, 1.2628], -fratioN[x, 2.616], -fratioP[x, 1.598]}
%

Meanwhile, we employ a phenomenological approach to the problem, kindred to those used, \emph{e.g}., in Refs.\,\cite{Liu:2019vsn, Han:2021dkc}, and exploit constraints suggested by analyses in perturbative QCD \cite{Brodsky:1994kg} to develop simple \emph{Ans\"atze} for the polarised DFs.
It is worth enumerating the constraints we impose.
\begin{enumerate}[label=(\emph{\alph*}), ref=(\alph*)]
\item At low-$x$, there is no correlation between the helicity of the struck quark and that of the parent proton; so, the polarised:unpolarised ratio of DFs must vanish: as $x \to 0$, $\Delta {\mathpzc q}(x;\zeta_{\cal H})/{\mathpzc q}(x;\zeta_{\cal H}) \to 0$.  Drawing on Regge phenomenology, we implement this by writing $\Delta {\mathpzc q}(x;\zeta_{\cal H}) \propto x^{\delta\alpha_R}{\mathpzc q}(x;\zeta_{\cal H})$, where $\delta\alpha_R = \tfrac{1}{2}$ is the difference between the intercepts of the vector and axialvector meson Regge trajectories \cite{Brisudova:1999ut}.
\item At high-$x$, the polarised and unpolarised valence quark distributions possess the same power-law behaviour, \emph{viz}.\ $\Delta {\mathpzc q}(x)/{\mathpzc q}(x) \to {\rm constant} \neq 0$ as $x\to 1$.
\end{enumerate}
These constraints are implemented using four distinct mappings: $\Delta {\mathpzc q}(x;\zeta_{\cal H}) \! = \! {\mathpzc s}_{\mathpzc q} {\mathpzc r}_i(x,\gamma_i^{\mathpzc q}) {\mathpzc q}(x;\zeta_{\cal H})$, with
${\mathpzc s}_{\mathpzc u} \! = \! 1 \! = \! - {\mathpzc s}_{\mathpzc d}$,
\begin{subequations}
\label{ratiofunctions}
\begin{align}
r_1(x,\gamma) & = \sqrt{x}/[1+\gamma \sqrt{x}] \,, &
r_2(x,\gamma) & = \sqrt{x}/[\gamma+ \sqrt{x}] \,, \\
r_3(x,\gamma) & = \sqrt{x}/[1+ \gamma x] \,, &
r_4(x,\gamma) & = \sqrt{x}/[\gamma+ x] \,.
\end{align}
\end{subequations}
%% fratioO[x_, \[Gamma]_] := Sqrt[x]/(1 + \[Gamma] Sqrt[x ])
%% fratio[x_, \[Gamma]_] := Sqrt[x]/(\[Gamma] + Sqrt[x ])
%% fratioN[x_, \[Gamma]_] := Sqrt[x]/(1 + \[Gamma] x)
%% fratioP[x_, \[Gamma]_] := Sqrt[x]/(\[Gamma] +  x)
In each case, $\gamma_i^{\mathpzc q}$ is fixed by requiring
%\begin{equation}
$\int_0^1 dx\,\Delta{\mathpzc q}(x;\zeta_{\cal H}) = g_A^{\mathpzc q}$.
%\end{equation}
Referring to Sec.\,\ref{PFE}-item~\ref{gAvalues}, this leads to the values listed in Table~\ref{gammavalues}.

\begin{figure}[t]
\centerline{\includegraphics[width=0.44\textwidth]{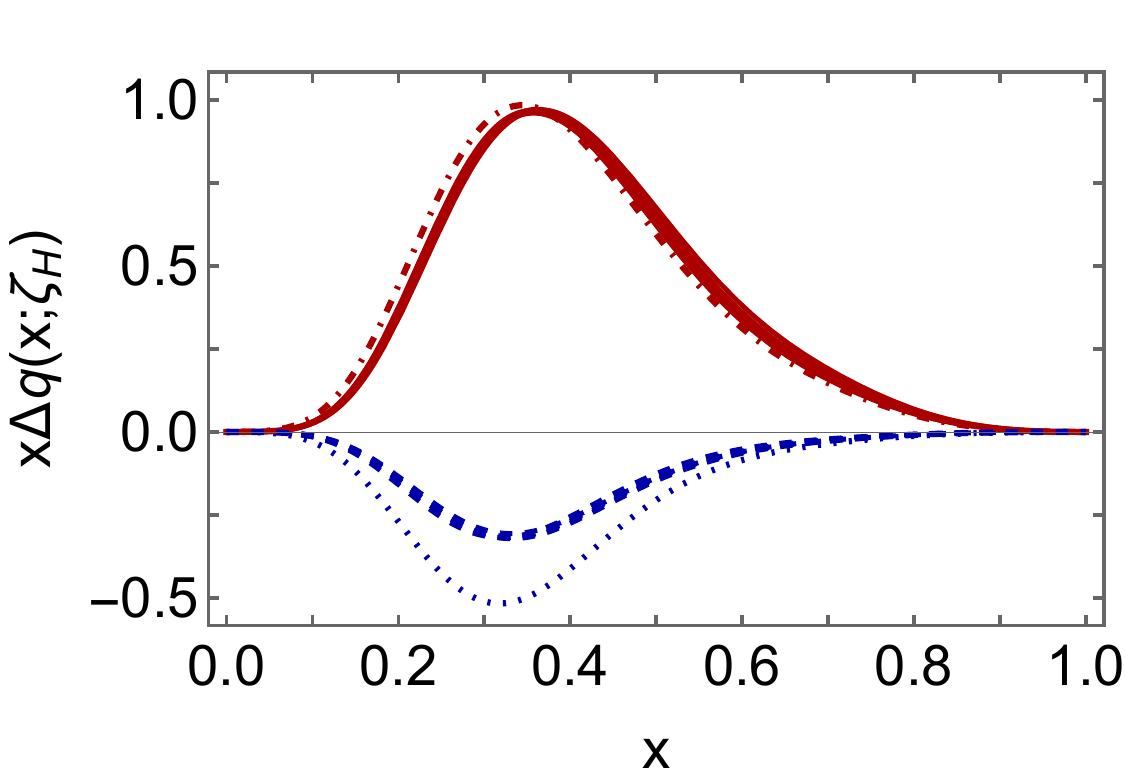}}
\caption{\label{FigPolarised}
Hadron scale polarised valence quark distributions: solid red curves -- $\mathpzc u$ quark; and dashed blue curves -- $\mathpzc d$ quark.  In each case, there are five curves, \emph{viz}.\ the four produced by the mapping functions in Eq.\,\eqref{ratiofunctions}, with the coefficient values in Table~\ref{gammavalues}, and the average of these curves.
Context is provided by the unpolarised valence quark distributions: $x{\mathpzc u}(x;\zeta_{\cal H})/2$ -- dot-dashed red curve; and $[-x{\mathpzc d}(x;\zeta_{\cal H})/2]$ -- dotted blue curve.
}
\end{figure}

Our \emph{Ans\"atze} for the hadron-scale polarised valence quark DFs are drawn in Fig.\,\ref{FigPolarised}, wherein they are compared with the kindred unpolarised DFs calculated in Ref.\,\cite{Chang:2022jri} -- reproduced by Eq.\,\eqref{ExpandGegenbauer} with the coefficients in Table~\ref{anValence}.

\begin{table}[t]
\begin{center}
\begin{tabular*}%{|c|c|c|c|c|c|c|}\hline
{\hsize}
{
l@{\extracolsep{0ptplus1fil}}
c@{\extracolsep{0ptplus1fil}}
c@{\extracolsep{0ptplus1fil}}
c@{\extracolsep{0ptplus1fil}}
c@{\extracolsep{0ptplus1fil}}
c@{\extracolsep{0ptplus1fil}}}\hline
 & herein & Faddeev & SC only & SU$(4)$ & pQCD \\\hline
$\rule{0ex}{3ex}\frac{F_2^n}{F_2^p}$ & $\phantom{-}0.45(5)$ & $\phantom{-}0.49$ & $\frac{1}{4}$ & $\phantom{-}\frac{2}{3}$ &$\frac{3}{7}$   \\[1ex]
$\frac{d}{u}$ & $\phantom{-}0.23(6)$ & $\phantom{-}0.28$ & 0 & $\phantom{-}\frac{1}{2}$ & $\frac{1}{5}$ \\[1ex]
$\frac{\Delta d}{\Delta u}$ & $-0.14(3)$ & $-0.11$ & 0 & $-\frac{1}{4}$ & $\frac{1}{5}$ \\[1ex]
$\frac{\Delta u}{u}$ &  $\phantom{-}0.63(8)$ & $\phantom{-}0.65$ & 1 & $\phantom{-}\frac{2}{3}$ & 1 \\[1ex]
$\frac{\Delta d}{d}$ & $-0.38(7)$ & $-0.26$ & 0 & $-\frac{1}{3}$ & 1 \\[1ex]
$A_1^n$ &  $\phantom{-}0.15(5)$ & $\phantom{-}0.17$ & 1 & $\phantom{-}0$ &  1 \\[1ex]
$A_1^p$ & $\phantom{-}0.58(8)$ & $\phantom{-}0.59$ & 1 & $\phantom{-}\frac{5}{9}$ & 1 \\\hline
\end{tabular*}
\caption{\label{x1Update}
Predictions for $x=1$ value of the indicated quantities.
Column ``herein'' collects predictions from Ref.\,\cite{Chang:2022jri} and those obtained using the polarised DFs in Fig.\,\ref{FigPolarised}.
``Faddeev'' reproduces the DSE-1/realistic results in Ref.\,\cite{Roberts:2013mja}, obtained using simple formulae, expressed in terms of diquark appearance and mixing probabilities.
The next two columns are, respectively, results drawn from Ref.\,\protect\cite{Close:1988br} -- proton modelled as being built using an elementary scalar diquark (no AV); and Ref.\,\cite{Hughes:1999wr} -- proton described by a SU$(4)$ spin-flavour wave function.
The last column, labelled ``pQCD,'' lists predictions made in Refs.\,\protect\cite{Farrar:1975yb, Brodsky:1994kg}, which assume an SU$(4)$ spin-flavour wave function for the proton's valence-quarks and that a hard photon may interact only with a quark that possesses the same helicity as the target.  ($3/7 \approx 0.43$.)
}
\end{center}
\end{table}

At this point, given that the $x=1$ value of any ratio of valence quark DFs is scale-independent \cite{Holt:2010vj}, we can provide an update of Ref.\,\cite[Table~1]{Roberts:2013mja} -- see Table~\ref{x1Update}.  The general agreement between our Faddeev equation based results and those in the ``Faddeev'' column indicates that reliable estimates are provided by the simple formulae introduced in Ref.\,\cite{Roberts:2013mja} for use in analysing nucleon Faddeev wave functions to obtain $x\to 1$ values of DF ratios without the need for calculating the $x$-dependence of any DF.  Viewed alternately, the agreement provides support for our polarised valence quark DF \emph{Ans\"atze}.

It is now appropriate to address the issue of helicity retention in hard scattering processes \cite{Farrar:1975yb, Brodsky:1994kg}.
If this notion is correct, then $\Delta d/d = 1=\Delta u/u$ on $x\simeq 1$ -- see Table~\ref{x1Update}-column~5.
However,
these ratios are invariant under QCD evolution (DGLAP \cite{Dokshitzer:1977sg, Gribov:1971zn, Lipatov:1974qm, Altarelli:1977zs});
such evolution cannot produce a zero in a valence quark DF;
and $\int_0^1 dx \Delta d(x;\zeta_{\cal H}) = g_A^d < 0$.
Consequently, helicity retention requires a zero in $\Delta d(x;\zeta_{\cal H})$.
Existing precision data indicate that if such a zero exists, then it must lie on $x\gtrsim 0.6$ \cite[HERMES]{HERMES:2004zsh}, \cite[COMPASS]{COMPASS:2010hwr},
\cite[CLAS EG1]{CLAS:2006ozz, CLAS:2008xos, CLAS:2015otq, CLAS:2017qga},
\cite[E06-014]{JeffersonLabHallA:2014mam},
\cite[E99-117]{JeffersonLabHallA:2003joy, JeffersonLabHallA:2004tea}.

Since we have modelled the polarised valence quark distributions, we cannot provide a CSM argument either in favour or against helicity retention.  In fact, no calculations of the polarised valence quark distributions are available in any nonperturbative framework with a traceable connection to QCD.  Nevertheless, the mappings in Eq.\,\eqref{ratiofunctions} preclude the possibility of a zero in $\Delta d(x;\zeta_{\cal H})$.  This choice is motivated by the observations that no viable direct calculation of $\Delta d(x;\zeta_{\cal H})$ delivers a result with a zero on the valence quark domain -- see, e.g, Refs.\,\cite{Deur:2018roz, Xu:2021wwj, Xu:2022abw}, and phenomenological DF global fits do not return a zero in $\Delta d(x)$ \cite{Ethier:2020way}.  Notwithstanding these things, a zero in $\Delta d(x;\zeta_{\cal H})$ is engineered in the model of Ref.\,\cite{Liu:2019vsn}.  All that may be said with certainty is that QCD-connected calculations of the polarised valence quark distributions are desirable as, too, are related data on $x \gtrsim 0.6$.  The latter exist \cite[CLAS RGC]{E1206109}, \cite[E12-06-110]{Zheng:2006} and completed analyses can reasonably be expected within a few years.
% Jefferson Lab Experiment E12-06-110 & CLAS EG12

A related issue concerns the $d/u$ (equivalently $F_2^n/F_2^p$) ratio in Table~\ref{x1Update}.
Columns ``herein'', ``Faddeev'' and ``pQCD'' agree, within uncertainties.
The first two are based on calculations of the proton's Poincar\'e-covariant wave function that include scalar and axialvector diquarks with dynamically prescribed relative strengths -- see Sec.\,\ref{PFE}-item\,\ref{item2i}.
Such a wave function corresponds to a structured leading-twist hadron-scale proton distribution amplitude (DA) \cite{Bali:2015ykx, Mezrag:2017znp}.
However, as the scale is increased, it is anticipated that all such structure is eliminated as the DA approaches its asymptotic form and the wave function comes to express SU$(4)$ spin-flavour symmetry \cite{Lepage:1980fj}.
In this case, given that $d/u(x=1)$ is invariant under evolution, then the ``pQCD'' prediction may be interpreted as a constraint on the relative strength of SC and AV correlations in the hadron-scale proton wave function; and that constraint is satisfied if, and only if, the Faddeev wave function has the properties described in Sec.\,\ref{PFE}-item\,\ref{item2i}.
Notably, this relative strength also provides an explanation \cite{Chang:2022jri, Lu:2022cjx} of modern data on $F_2^n(x)/F_2^p(x)$ \cite[MARATHON]{Abrams:2021xum} and its extrapolation \cite{Cui:2021gzg}: on $x\simeq 1$, $F_2^n/F_2^p = 0.437(85)$.

\emph{Caveat~1}.  As a prelude to continuing, it is important to observe that all polarisation ``data'' reproduced herein were obtained from analyses of experiments that employ one or another set of the available global DF fits to another body of experiments.  The ``data'' values and uncertainties are therefore contingent upon the reliability of the chosen global fit.  Furthermore, there is no guarantee of consistency between the given new experiment and the body used to produce the existing global fit.  Consequently, the reported ``data'' are not objective.  In contrast, as we now explain, the predictions made herein follow from an internally consistent, unified treatment of all DFs.

\begin{figure}[t]
\centerline{\includegraphics[width=0.45\textwidth]{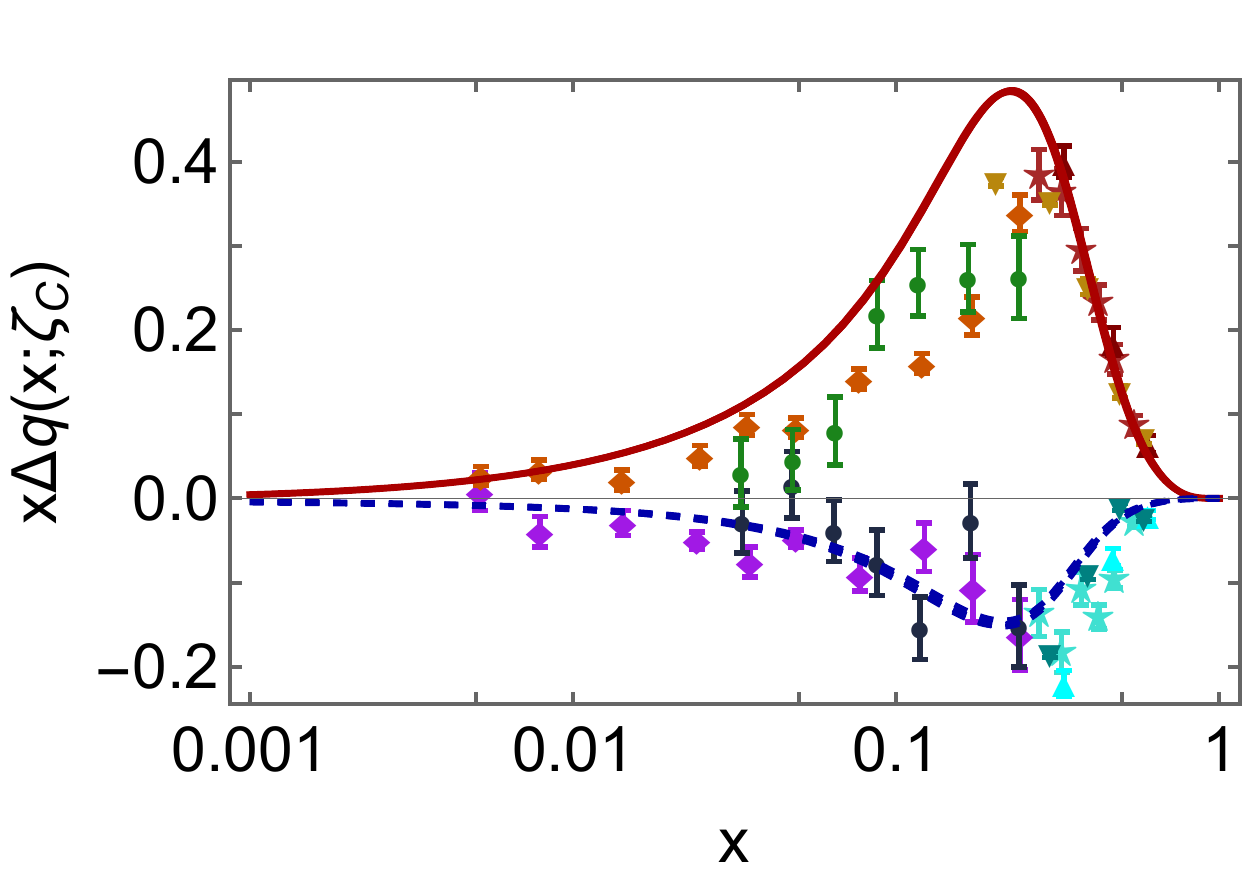}}
\caption{\label{polzetaCa}
Polarised quark DFs: $\Delta {\mathpzc u}(x;\zeta_{\rm C})$ -- solid red curves; and $\Delta {\mathpzc d}(x;\zeta_{\rm C})$ -- dashed blue curves.
Data: \cite[HERMES]{HERMES:2004zsh} -- circles; \cite[COMPASS]{COMPASS:2010hwr} -- diamonds;
filled down-triangles -- \cite[CLAS EG1]{CLAS:2006ozz, CLAS:2008xos, CLAS:2015otq, CLAS:2017qga};
five-pointed stars -- \cite[E06-014]{JeffersonLabHallA:2014mam};
filled up-triangles --\cite[E99-117]{JeffersonLabHallA:2003joy, JeffersonLabHallA:2004tea}.
}
\end{figure}

\begin{figure}[t]
%\vspace*{6ex}

\leftline{\hspace*{0.5em}{{\textsf{(a)}}}}
\vspace*{-4ex}
\hspace*{1em}\includegraphics[width=0.44\textwidth]{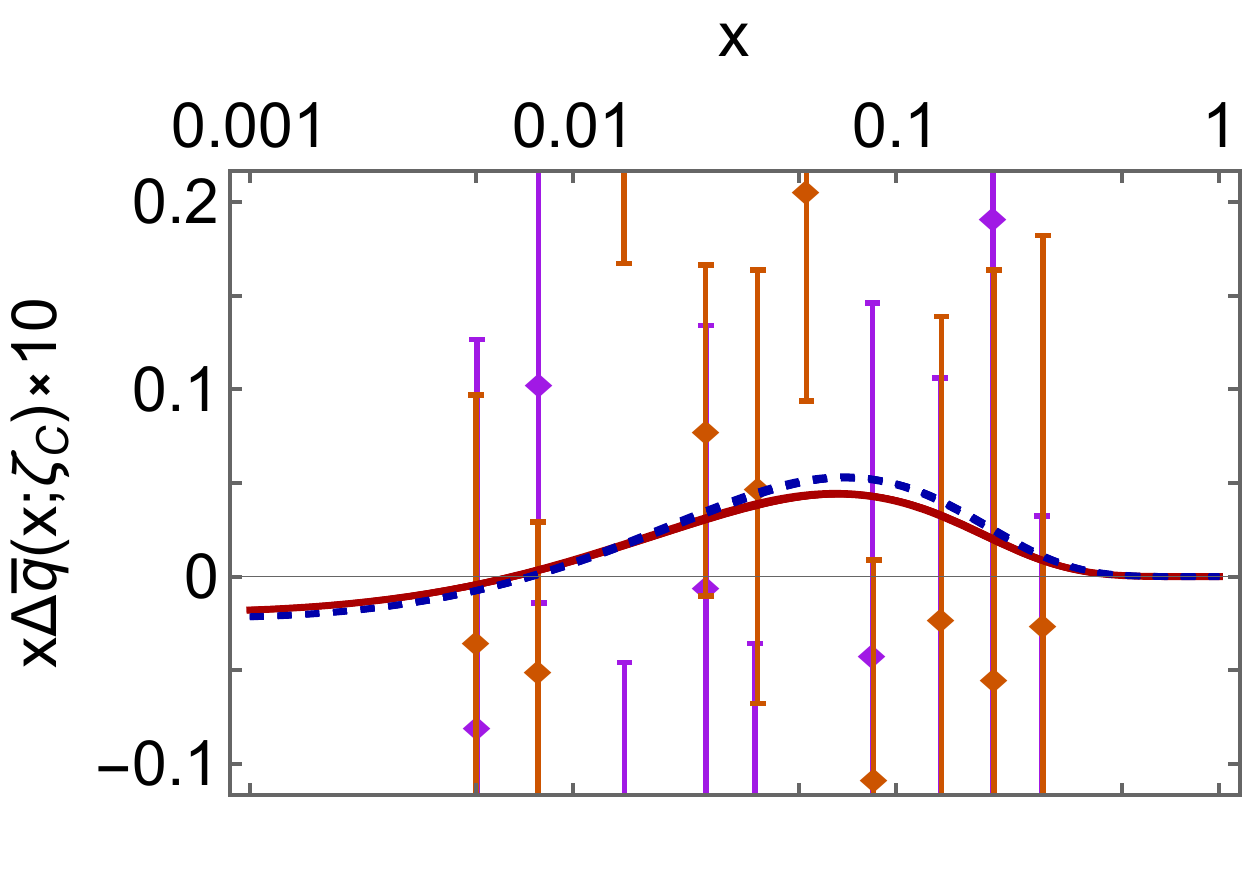}
\vspace*{-3ex}

\leftline{\hspace*{0.5em}{{\textsf{(b)}}}}
\vspace*{-4ex}
\hspace*{1em}\includegraphics[width=0.44\textwidth]{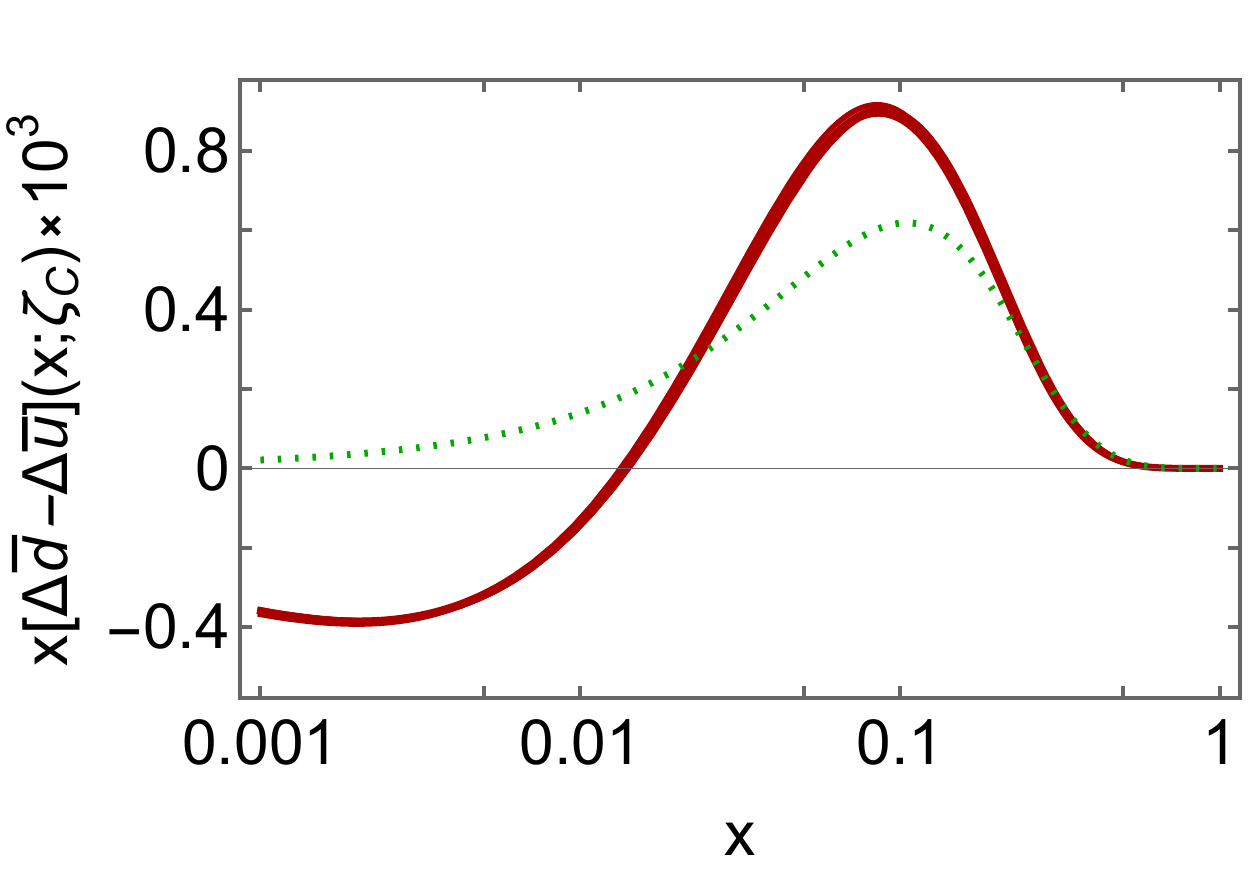}

\caption{\label{polzetaCB}
\textsf{(a)}
Polarised antiquark DFs: $\Delta \bar{\mathpzc u}(x;\zeta_{\rm C})$ -- solid red curves; and $\Delta \bar{\mathpzc d}(x;\zeta_{\rm C})$ -- dashed blue curves.
Data \cite[COMPASS]{COMPASS:2010hwr}: $\bar{\mathpzc u}$ -- red diamonds; and $\bar{\mathpzc d}$ -- purple squares.
\textsf{(b)}
$x[\Delta \bar{\mathpzc d}(x;\zeta_{\rm C})-\Delta \bar{\mathpzc u}(x;\zeta_{\rm C})]$ -- solid red curves.
For comparison: $x[\bar{\mathpzc d}(x;\zeta_{\rm C})- \bar{\mathpzc u}(x;\zeta_{\rm C})]$ -- dotted green curve.
}
\end{figure}

\section{Polarised quark distributions at $\zeta^2=3\,$GeV$^2$}
With $\zeta_{\cal H}$ being that scale at which all properties of the proton are carried by its dressed valence quark dof, then the all-orders extension of QCD evolution explained in Refs.\,\cite{Cui:2019dwv, Raya:2021zrz, Cui:2021mom, Cui:2022bxn} can be used to obtain all proton DFs at any scale $\zeta > \zeta_{\cal H}$.  This approach has been used with success to predict proton and pion unpolarised DFs \cite{Lu:2022cjx} and, recently, to extract the pion mass distribution from available data \cite{Xu:2023bwv}.  Herein, we use it to deliver predictions for proton polarised DFs.

This all-orders evolution scheme is based on a single proposition \cite{Raya:2021zrz, Cui:2021mom, Cui:2022bxn}:\\
\hspace*{0.75em}\parbox[t]{0.95\linewidth}{{\sf P1} -- \emph{In the context of Refs.\,\cite{Grunberg:1980ja, Grunberg:1982fw}, there exists at least one effective charge, $\alpha_{1\ell}(k^2)$, which, when used to integrate the leading-order perturbative DGLAP equations, defines an evolution scheme for parton DFs that is all-orders exact}.}  %$\alpha_{1\ell}(k^2)$ need not be unique.
\smallskip

\noindent Such charges  need not be process-independent (PI); hence, not unique.
%: different charges may be needed for distinct observables.
Nevertheless, an efficacious PI charge is not excluded.  That discussed in Refs.\,\cite{Cui:2019dwv, Cui:2020tdf, Raya:2021zrz}, denoted $\hat\alpha$, has proved suitable and we employ it herein.  Using $\hat\alpha$, one predicts $\zeta_{\cal H}=0.331(2)\,$GeV.
Connections with experiment and other nonperturbative extensions of QCD's running coupling are given in Refs.\,\cite{Deur:2016tte, Deur:2022msf, Deur:2023dzc}.

Implemented as described in Ref.\,\cite{Lu:2022cjx} and applied to the polarised valence DFs in Fig.\,\ref{FigPolarised}, one obtains the $\zeta_{\rm C}=\surd 3\,$GeV polarised quark DFs drawn in Fig.\,\ref{polzetaCa}, wherein they are compared with data inferred from experiments \cite[HERMES]{HERMES:2004zsh}, \cite[COMPASS]{COMPASS:2010hwr},
\cite[CLAS EG1]{CLAS:2006ozz, CLAS:2008xos, CLAS:2015otq, CLAS:2017qga},
\cite[E06-014]{JeffersonLabHallA:2014mam},
\cite[E99-117]{JeffersonLabHallA:2003joy, JeffersonLabHallA:2004tea}:
there is agreement on the valence quark domain, $x\gtrsim 0.2$.

\begin{figure}[t]
\centerline{\includegraphics[width=0.44\textwidth]{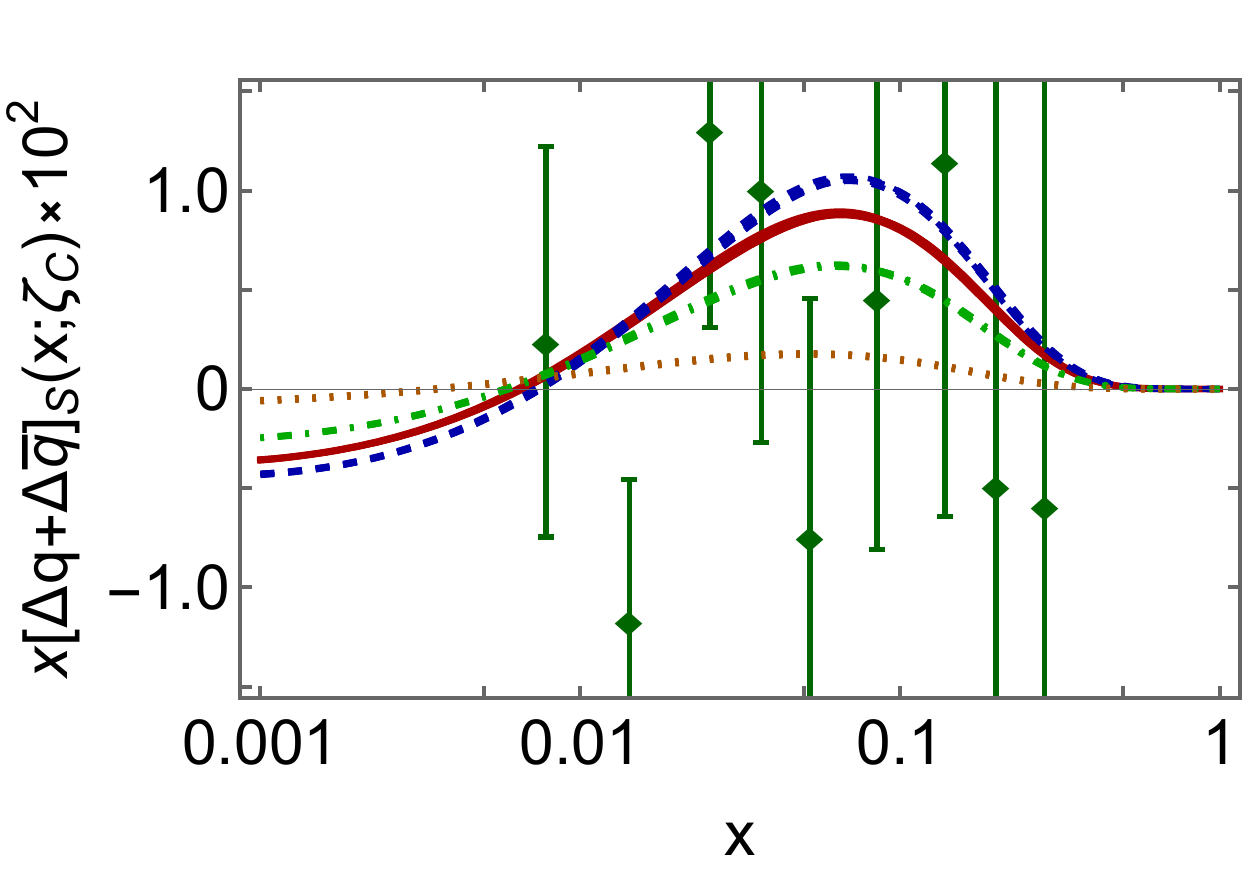}}
\caption{\label{FigPolarisedSea}
Polarised sea quark distributions at $\zeta_{\rm C}$.
Solid red curves: $x[\Delta{\mathpzc u}+\Delta\bar{\mathpzc u}]_S$; dashed blue curves: $x[\Delta{\mathpzc d}+\Delta\bar{\mathpzc d}]_S$; $x[\Delta{\mathpzc s}+\Delta\bar{\mathpzc s}]_S$: dot-dashed green curves; $x[\Delta{\mathpzc c}+\Delta\bar{\mathpzc c}]_S$: dotted orange curves.
Context is provided by values of $2x\Delta {\mathpzc s}(x;\zeta_{\rm C})$ from Ref.\cite[COMPASS]{COMPASS:2010hwr}, wherein $x\Delta {\mathpzc s}(x;\zeta_{\rm C}) \approx x\Delta \bar {\mathpzc s}(x;\zeta_{\rm C})$.
}
\end{figure}

Referring to the COMPASS results, lying on $x\lesssim 0.2$, the collaboration's extrapolations yield $g_A^d=-0.34(5)$, $g_A^u=0.71(4)$, $g_A=g_A^u-g_A^d=1.05(6)$.
Comparison with Sec.\,\ref{PFE}-item\,\ref{gAvalues}, shows agreement with our value of $g_A^d$.  This is reflected in the match between data and our prediction for $\Delta {\mathpzc d}(x;\zeta_{\rm C})$.
On the other hand, the data tend to lie below our prediction for $\Delta {\mathpzc u}(x;\zeta_{\rm C})$.
This is understandable.  The COMPASS results lead to a value for $g_A^u-g_A^d$ that is too small, \emph{viz}.\ $0.83(5)$-times the value determined from neutron $\beta$-decay, an outcome which can be attributed to a low value of $g_A^u$: it is only $0.75(3)$-times our prediction.  (These statements can be made because polarised antiquark DFs are negligible at this scale -- see Figs.\,\ref{polzetaCB}, \ref{FigPolarisedSea}.)

The polarised antiquark distributions are drawn in Fig.\,\ref{polzetaCB}a along with values from Ref.\,\cite[COMPASS]{COMPASS:2010hwr}.
On the scale of this image, set by the magnitudes of our predictions, the data have large uncertainties; so, can only be used to set plausible bounds on the size of these distributions.

The difference $x[\Delta \bar{\mathpzc d}(x;\zeta_{\rm C})-\Delta \bar{\mathpzc u}(x;\zeta_{\rm C})]$ is depicted in Fig.\,\ref{polzetaCB}b and compared with the result for $x[\bar{\mathpzc d}(x;\zeta_{\rm C})- \bar{\mathpzc u}(x;\zeta_{\rm C})]$ from Refs.\,\cite{Chang:2022jri, Lu:2022cjx}, which reproduce the proton antimatter asymmetry reported in Ref.\,\cite[SeaQuest]{SeaQuest:2021zxb}.  (This was achieved via a modest Pauli blocking factor in the gluon splitting function.)
Notably, both differences have the same magnitude; and their trend is similar on $x\gtrsim 0.01$: using {\sf P1}, they are related.
In this case, we do not report a comparison with data because the uncertainties on available results \cite[HERMES]{HERMES:2004zsh}, \cite[COMPASS]{COMPASS:2010hwr} are too large for such a comparison to be meaningful.

Our predictions for all polarised sea quark DFs are collected in Fig.\,\ref{FigPolarisedSea}.  Following the implementation of all-orders evolution in Ref.\,\cite{Lu:2022cjx}, we have thresholds at which heavier quarks begin to play a role in evolution.  This explains the flavour-separation amongst the polarised sea DFs.
The image includes results on $2x\Delta {\mathpzc s}(x;\zeta_{\rm C})$ inferred from data in Ref.\cite[COMPASS]{COMPASS:2010hwr}.  Broadly speaking, the magnitude matches our prediction for this DF; but, again, the data uncertainties are large.  Our prediction
$\int_{0.004}^{0.3} dx \,\Delta{\mathpzc s}_S(x;\zeta_{\rm C}) = 0.0072(1)$ is consistent with the inferred empirical value \cite[COMPASS]{COMPASS:2010hwr}: $-0.01\pm 0.01\pm 0.01$.
%% 0.00722538 +/- 0.0000993116

\begin{figure}[t]
\centerline{\includegraphics[width=0.44\textwidth]{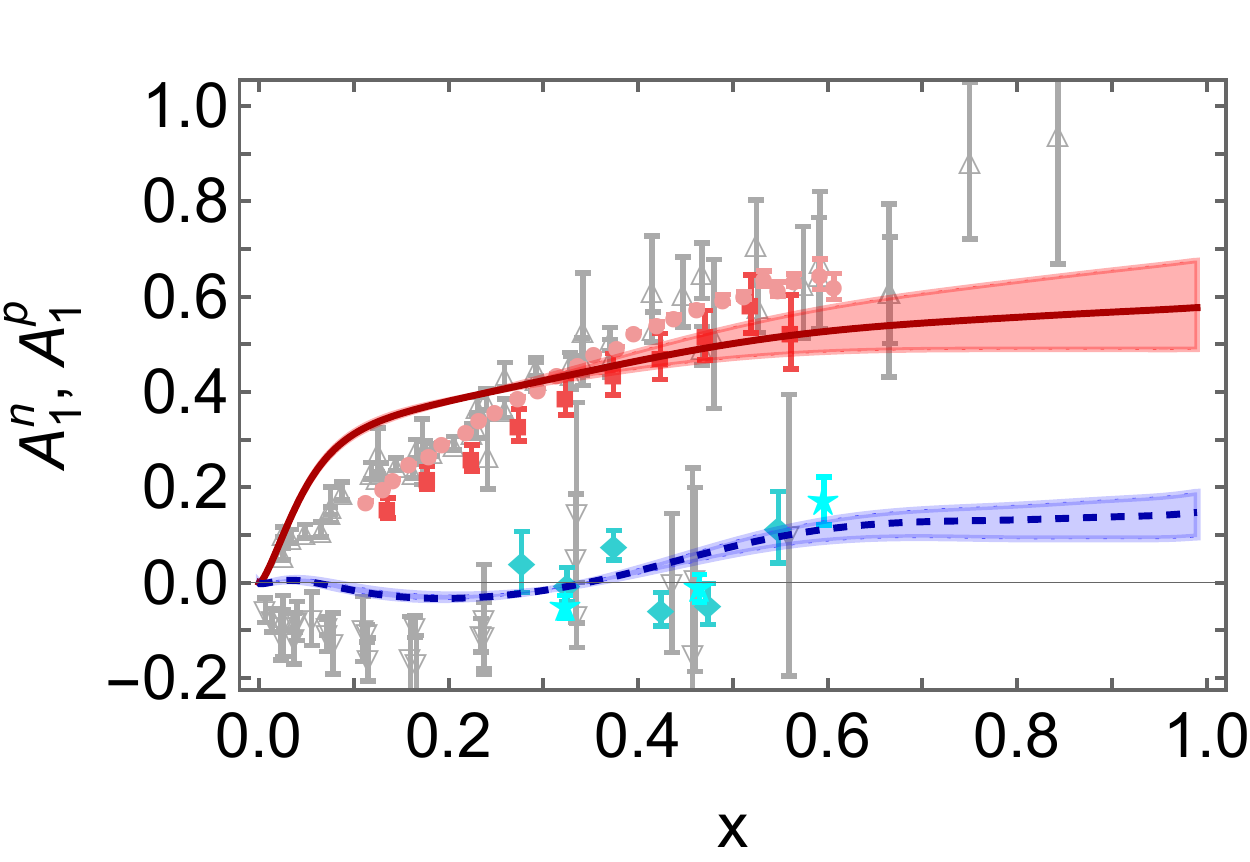}}
\caption{\label{FigA1}
Predictions for proton (solid red) and neutron (dashed blue) longitudinal spin asymmetries at $\zeta_{\rm C}$.  In each case, the associated band expresses the uncertainty deriving from Eq.\,\eqref{ratiofunctions}.
Data.
$A_1^p$:
red squares -- Refs.\,\cite[CLAS EG1]{CLAS:2006ozz, CLAS:2008xos, CLAS:2015otq, CLAS:2017qga};
pink circles -- Ref.\,\cite{CLAS:2014qtg};
grey up-triangles -- Refs.\,\cite{E155:1999pwm, E155:2000qdr, HERMES:1998cbu, E143:1995clm, E143:1996vck, E143:1998hbs, SpinMuonSMC:1994met, SpinMuonSMC:1997mkb}.
$A_1^n$: turquoise diamonds -- Ref.\,\cite[E06-014]{JeffersonLabHallA:2014mam};
aqua five-pointed stars -- Refs.\,\cite[E99-117]{JeffersonLabHallA:2003joy, JeffersonLabHallA:2004tea};
grey down-triangles -- Refs.\,\cite{HERMES:1997hjr, E154:1997xfa, E154:1997ysl, E142:1996thl}.
}
\end{figure}

In Fig.\,\ref{FigA1}, we depict our predictions for nucleon longitudinal spin asymmetries -- defined, \emph{e.g}., as in Ref.\,\cite[Ch.\,4.7]{Ellis:1991qj}.  (The $c$ quark contribution is practically negligible at this scale.)
For context, we also show results inferred from data collected within the past vicennium \cite{CLAS:2006ozz, CLAS:2008xos, CLAS:2015otq, CLAS:2017qga, CLAS:2014qtg, JeffersonLabHallA:2014mam, JeffersonLabHallA:2003joy, JeffersonLabHallA:2004tea} and selected earlier results \cite{E155:1999pwm, E155:2000qdr, HERMES:1998cbu, E143:1995clm, E143:1996vck, E143:1998hbs, SpinMuonSMC:1994met, SpinMuonSMC:1997mkb, HERMES:1997hjr, E154:1997xfa, E154:1997ysl, E142:1996thl}.
The mismatch between prediction and inferences at low-$x$ may reflect known discrepancies between our predictions for sea quark DFs and those produced by phenomenological fits \cite{Cui:2020tdf, Lu:2022cjx}.  On the other hand, there is general agreement between our predictions and data on $x\gtrsim 0.2$.  New experiments able to return DF information on $x\gtrsim 0.6$ are desirable; especially in connection with the question of helicity retention discussed earlier.  Thus, analyses of data collected recently \cite[CLAS RGC]{E1206109}, \cite[E12-06-110]{Zheng:2006} are much anticipated.

\section{Polarised gluon distribution at $\zeta^2=3\,$GeV$^2$}
\begin{figure}[t]
%\vspace*{6ex}

\leftline{\hspace*{0.5em}{{\textsf{(a)}}}}
\vspace*{-4ex}
\hspace*{1em}\includegraphics[width=0.44\textwidth]{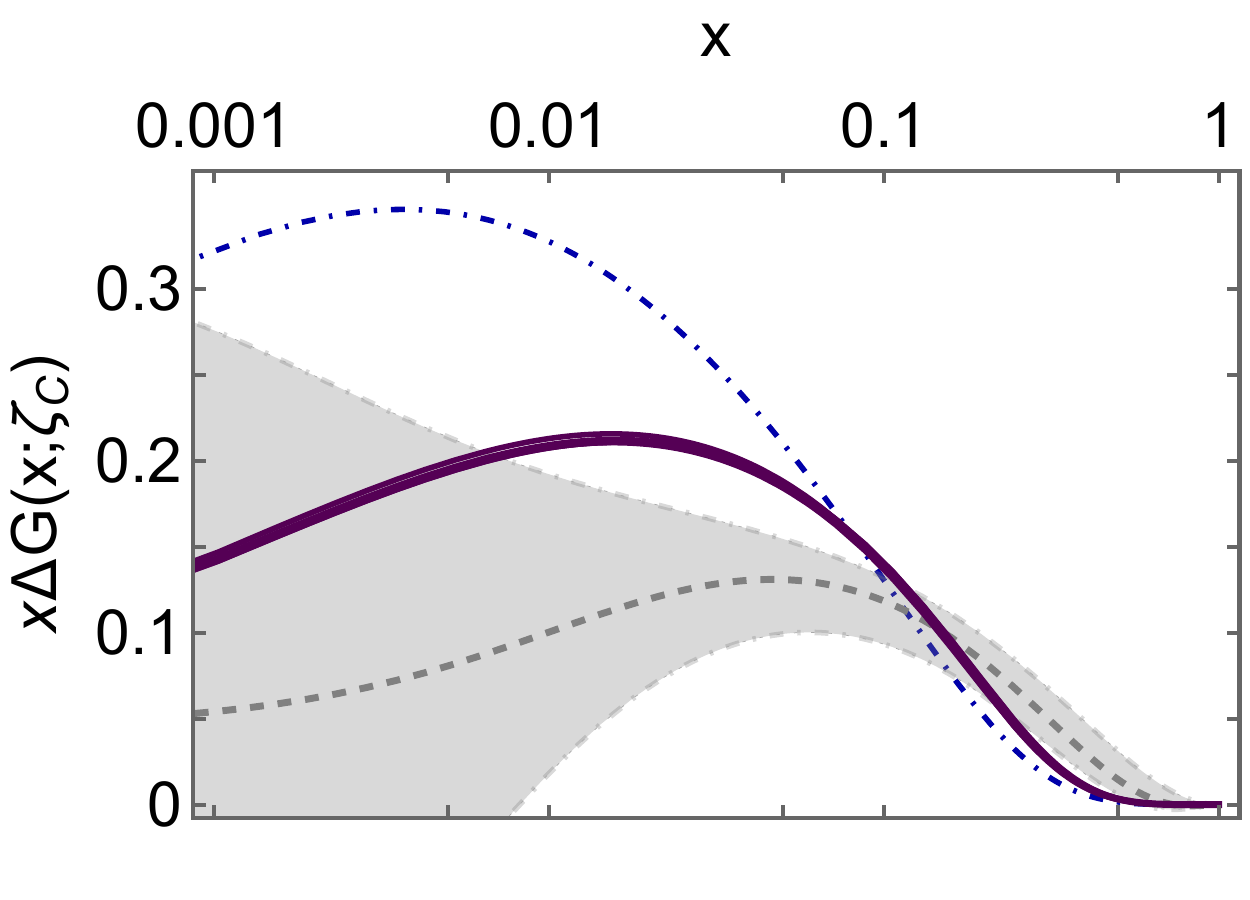}
\vspace*{-3ex}

\leftline{\hspace*{0.5em}{{\textsf{(b)}}}}
\vspace*{-4ex}
\hspace*{1em}\includegraphics[width=0.44\textwidth]{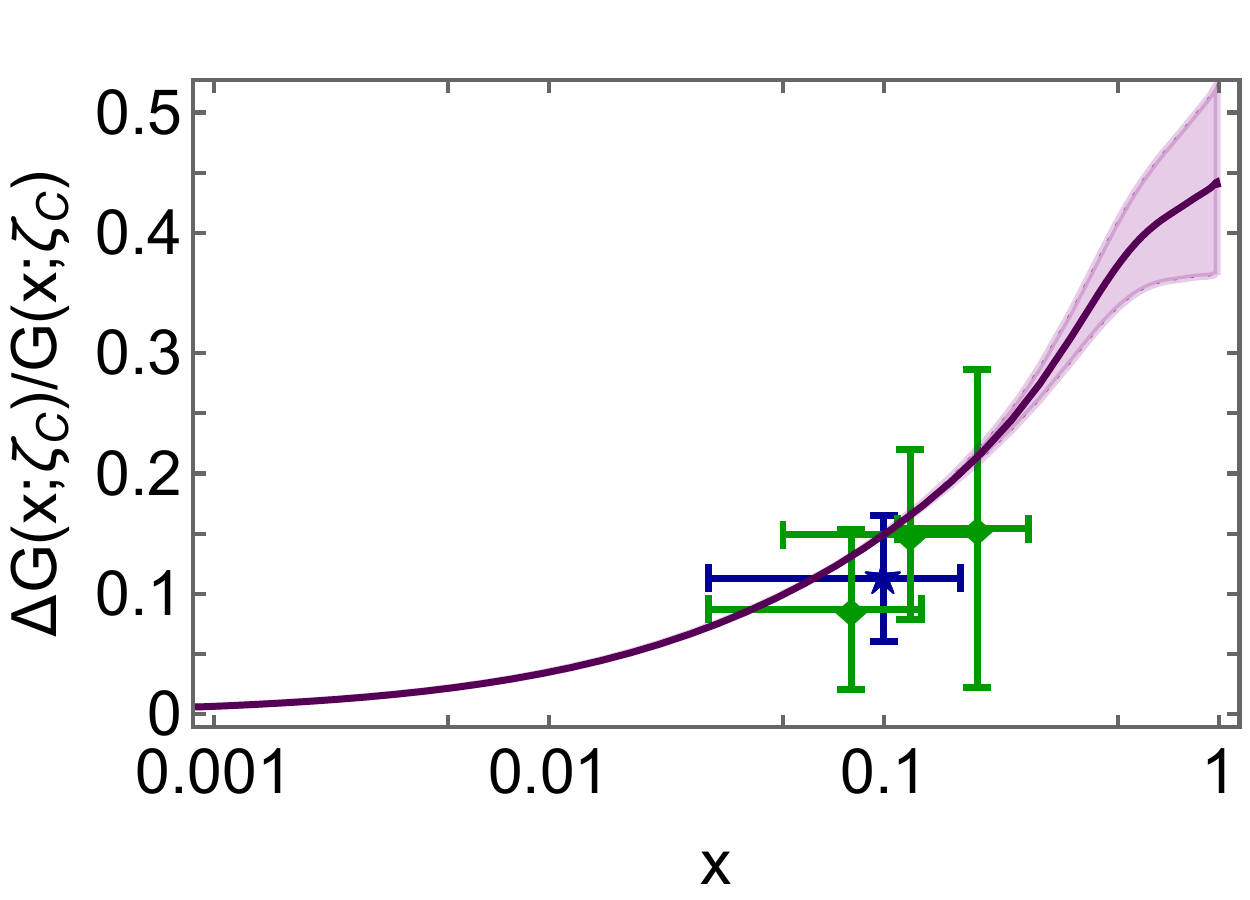}

\caption{\label{Figgluon}
\textsf{(a)}
Polarised gluon DF: $\Delta G(x;\zeta_{\rm C})$ -- solid purple curves.
Gray dashed curve and band: typical result from phenomenological global fit \cite[DSSV14]{deFlorian:2014yva} reported at $\zeta=10\,$GeV.
Evolved to this scale, our central prediction is the blue dot-dashed curve.
\textsf{(b)}
Polarised/unpolarised DF ratio $\Delta G(x;\zeta_{\rm C})/G(x;\zeta_{\rm C})$.  The associated band expresses the uncertainty deriving from Eq.\,\eqref{ratiofunctions}.
For context, we depict values reported in Ref.\,\cite[COMPASS]{COMPASS:2015pim}.
}
\end{figure}

Beginning with the hadron-scale DFs in Fig.\,\ref{FigPolarised}, the all-orders evolution scheme delivers the polarised and unpolarised gluon DFs at any scale $\zeta>\zeta_{\cal H}$.  Our $\zeta_{\rm C}$ predictions are drawn in Fig.\,\ref{Figgluon}.

Insofar as phenomenological DF fits are concerned, $\Delta G(x)$ is very poorly constrained.   This is illustrated by the grey band in Fig.\,\ref{Figgluon}a, drawn from Ref.\,\cite[DSSV14]{deFlorian:2014yva} at $\zeta=10\,$GeV.  At this scale, our central result is indicated by the dot-dashed (blue) curve.  Evidently, as also seen with unpolarised DFs, on $x\lesssim 0.05$, our internally consistent predictions for glue (and sea) DFs are larger in magnitude than those inferred through phenomenological fits \cite{Cui:2020tdf, Lu:2022cjx}.  Notwithstanding this, on the complementary domain we find
\begin{equation}
\int_{0.05}^1 dx \, \Delta G(x;\zeta=10\,{\rm GeV}) = 0.199(3)\,,
\end{equation}
\emph{cf}.\, $0.19(6)$ in Ref.\,\cite[DSSV14]{deFlorian:2014yva}.

Our prediction for the ratio $\Delta G(x;\zeta_{\rm C})/G(x;\zeta_{\rm C})$ is drawn in Fig.\,\ref{Figgluon}b.  There is good agreement with the results reported in Ref.\,\cite[COMPASS]{COMPASS:2015pim}.  This is highlighted, \emph{e.g}., by noting that the mean value of our result on the domain covered by measurements is $0.167(3)$ \emph{cf}.\, $ 0.113 \pm 0.038 \pm 0.036$ \cite[COMPASS]{COMPASS:2015pim}.

\section{Proton spin}
It is now appropriate to recall Eq.\,\eqref{isoscalaraxialcharge}, which records that $\approx 65$\% of the proton spin is carried by valence quark quasiparticle degrees of freedom at the hadron scale.  Under {\sf P1}, this value is independent of scale.

On the other hand, measurements of the proton spin are sensitive to the non-Abelian anomaly corrected combination \cite{Altarelli:1988nr}
\begin{equation}
\label{Eqa0E}
a_0^{\rm E}(\zeta) = a_0 - n_f \frac{\hat \alpha(\zeta)}{2\pi} \int_0^1 dx\,\Delta G(x;\zeta)
=: a_0 - n_f \frac{\hat \alpha(\zeta)}{2\pi} \Delta G(\zeta)\,,
\end{equation}
where $n_f$ is the number of active quark flavours: herein, $\hat\alpha$ and {\sf P1} evolution are defined using $n_f=4$.

Using our result for $\Delta G(x;\zeta_{\rm C})$ -- Fig.\,\ref{Figgluon} -- to evaluate the right-hand side of Eq.\,\eqref{Eqa0E}, we predict
\begin{equation}
a_0^{\rm E}(\zeta_{\rm C}) = 0.35(2)\,.
\end{equation}
%% = 0.32±0.02stat ±0.04syst ±0.05evol .
This value compares favourably with that reported in Ref.\,\cite[COMPASS]{COMPASS:2016jwv}: $0.32(7)$.

Notwithstanding the remarks in the Introduction, it is common to report a light-front breakdown of the proton spin into contributions from quark and gluon spin and angular momenta:
\begin{equation}
\frac{1}{2} = \frac{1}{2} a_0 + L_q(\zeta)+\Delta G(\zeta)+L_g(\zeta)\,.
\end{equation}
% where $\Delta G(\zeta) = \int_0^1 dx\,\Delta G(x;\zeta)$.
This can be accomplished without ambiguity by exploiting the character of the hadron scale, \emph{viz}.\
\begin{equation}
L_q(\zeta_{\cal H}) = \frac{1}{2} - \frac{1}{2} a_0 = 0.175\,,
\Delta G(\zeta_{\cal H})=0=L_g(\zeta_{\cal H})\,,
\end{equation}
and subsequently employing {\sf P1} evolution, which is implemented with minor modifications of Eqs.\,(32) in Ref.\,\cite{Deur:2018roz}.  In this way, we determine the following central values:
\begin{equation}
L_q(\zeta_{\rm C}) = -0.027\,,\;
\Delta G(\zeta_{\rm C}) = 1.23\,,\;
L_g(\zeta_{\rm C}) =-1.03\,.
\end{equation}
Evidently, although it begins positive, the light-front quark angular momentum fraction falls steadily with increasing scale; and the increasing gluon helicity is compensated by a growth in magnitude of the light-front gluon angular momentum fraction.  The asymptotic ($\zeta\to\infty$) limits are discussed elsewhere \cite{Ji:1995cu, Chen:2011gn}.

\section{Summary and Outlook}
Beginning with \emph{Ans\"atze} for proton hadron-scale polarised valence quark distribution functions, developed using insights from perturbative QCD and constrained by solutions of a quark+diquark Faddeev equation, and supposing that there is an effective charge which defines an evolution scheme for parton DFs -- both unpolarised and polarised -- that is all-orders exact, we delivered parameter-free predictions for all proton polarised DFs at the scale $\zeta_{\rm C}^2 = 3\,$GeV$^2$.  In doing so, we completed a unification of proton and pion DFs.  All predictions, both pointwise behaviour and moments, compare favourably with results inferred from data, as exemplified herein by our results for polarised quark DFs [Fig.\,\ref{polzetaCa}] and nucleon longitudinal spin asymmetries [Fig.\,\ref{FigA1}].

Of particular significance is our finding that the polarised gluon DF, $\Delta G(x;\zeta_{\rm C})$, in the proton is positive and large [Fig.\,\ref{Figgluon}].  This prediction can be tested using experiments at next-generation QCD facilities \cite{Anderle:2021wcy, AbdulKhalek:2021gbh}.  Meanwhile, our result for $\Delta G(x;\zeta_{\rm C})$ enables us to predict that measurements of the proton singlet axial charge should return a value $a_0^{\rm E}(\zeta_{\rm C}) = 0.35(2)$.  This result is in accord with contemporary data.

Our analysis can be tested and improved in future by using a recently developed symmetry preserving axial current, appropriate for a proton described as a bound state of dressed-quark and fully interacting nonpointlike diquark degrees-of-freedom, to calculate the proton hadron scale polarised valence quark DFs.  A first step in this direction is underway using a careful treatment of a momentum-independent quark+quark interaction.  Its extension to a study of the proton using QCD-connected Schwinger functions is thereafter natural.  Longer term goals include analogous calculations that begin with a Poincar\'e-covariant three-body treatment of the nucleon bound state problem \cite{Eichmann:2009qa, Wang:2018kto}.

%\section*{Acknowledgments}
\medskip
\noindent\textbf{Acknowledgments}.
We are grateful for assistance and constructive comments from
A.~Deur,
T.~Liu
and
C.~Mondal.
%O.~Denisov, J.~Friedrich, W.-D.~Nowak, C.~Quintans and S.~Platchkov.
%
Work supported by:
National Natural Science Foundation of China (grant nos.\,12135007, 12247103);
Natural Science Foundation of Jiangsu Province (grant no.\ BK20220323).

%%\medskip
%%\noindent\textbf{Data Availability Statement}. This manuscript has no associated data or the data will not be deposited. [Authors' comment: All information necessary to reproduce the results described herein is contained in the material presented above.]

\medskip
\noindent\textbf{Declaration of Competing Interest}.
The authors declare that they have no known competing financial interests or personal relationships that could have appeared to influence the work reported in this paper.

%\section*{References}

%%\bibliographystyle{elsarticle-num-names}
%%\bibliography{../../../CollectedBiB}

\begin{thebibliography}{100}
\providecommand{\natexlab}[1]{#1}
\providecommand{\url}[1]{\texttt{#1}}
\providecommand{\urlprefix}{URL }
\expandafter\ifx\csname urlstyle\endcsname\relax
  \providecommand{\doi}[1]{doi:\discretionary{}{}{}#1}\else
  \providecommand{\doi}[1]{doi:\discretionary{}{}{}\begingroup
  \urlstyle{rm}\url{#1}\endgroup}\fi
\providecommand{\bibinfo}[2]{#2}

\bibitem[{Brodsky et~al.(2022)Brodsky, Deur, and Roberts}]{Brodsky:2022fqy}
\bibinfo{author}{S.~J. Brodsky}, \bibinfo{author}{A.~Deur},
  \bibinfo{author}{C.~D. Roberts}, \bibinfo{title}{{Artificial dynamical
  effects in quantum field theory}}, \bibinfo{journal}{Nature Rev. Phys.}
  \bibinfo{volume}{4}~(\bibinfo{number}{7}) (\bibinfo{year}{2022})
  \bibinfo{pages}{489--495}.

\bibitem[{Eichmann et~al.(2010)Eichmann, Alkofer, Krassnigg, and
  Nicmorus}]{Eichmann:2009qa}
\bibinfo{author}{G.~Eichmann}, \bibinfo{author}{R.~Alkofer},
  \bibinfo{author}{A.~Krassnigg}, \bibinfo{author}{D.~Nicmorus},
  \bibinfo{title}{{Nucleon mass from a covariant three-quark Faddeev
  equation}}, \bibinfo{journal}{Phys. Rev. Lett.} \bibinfo{volume}{104}
  (\bibinfo{year}{2010}) \bibinfo{pages}{201601}.

\bibitem[{Wang et~al.(2018)Wang, Qin, Roberts, and Schmidt}]{Wang:2018kto}
\bibinfo{author}{Q.-W. Wang}, \bibinfo{author}{S.-X. Qin},
  \bibinfo{author}{C.~D. Roberts}, \bibinfo{author}{S.~M. Schmidt},
  \bibinfo{title}{{Proton tensor charges from a Poincar{\'e}-covariant Faddeev
  equation}}, \bibinfo{journal}{Phys. Rev. D} \bibinfo{volume}{98}
  (\bibinfo{year}{2018}) \bibinfo{pages}{054019}.

\bibitem[{Ashman et~al.(1988)}]{EuropeanMuon:1987isl}
\bibinfo{author}{J.~Ashman}, et~al., \bibinfo{title}{{A Measurement of the Spin
  Asymmetry and Determination of the Structure Function $g_1$ in Deep Inelastic
  Muon-Proton Scattering}}, \bibinfo{journal}{Phys. Lett. B}
  \bibinfo{volume}{206} (\bibinfo{year}{1988}) \bibinfo{pages}{364}.

\bibitem[{Brodsky and Lepage(1989)}]{Brodsky:1989pv}
\bibinfo{author}{S.~J. Brodsky}, \bibinfo{author}{G.~P. Lepage},
  \bibinfo{title}{{Exclusive Processes in Quantum Chromodynamics}},
  \bibinfo{journal}{Adv. Ser. Direct. High Energy Phys.} \bibinfo{volume}{5}
  (\bibinfo{year}{1989}) \bibinfo{pages}{93--240}.

\bibitem[{Brodsky et~al.(1998)Brodsky, Pauli, and Pinsky}]{Brodsky:1997de}
\bibinfo{author}{S.~J. Brodsky}, \bibinfo{author}{H.-C. Pauli},
  \bibinfo{author}{S.~S. Pinsky}, \bibinfo{title}{{Quantum chromodynamics and
  other field theories on the light cone}}, \bibinfo{journal}{Phys. Rept.}
  \bibinfo{volume}{301} (\bibinfo{year}{1998}) \bibinfo{pages}{299--486}.

\bibitem[{Heinzl(2001)}]{Heinzl:2000ht}
\bibinfo{author}{T.~Heinzl}, \bibinfo{title}{{Light cone quantization:
  Foundations and applications}}, \bibinfo{journal}{Lect. Notes Phys.}
  \bibinfo{volume}{572} (\bibinfo{year}{2001}) \bibinfo{pages}{55--142}.

\bibitem[{Eichmann et~al.(2016)Eichmann, Sanchis-Alepuz, Williams, Alkofer, and
  Fischer}]{Eichmann:2016yit}
\bibinfo{author}{G.~Eichmann}, \bibinfo{author}{H.~Sanchis-Alepuz},
  \bibinfo{author}{R.~Williams}, \bibinfo{author}{R.~Alkofer},
  \bibinfo{author}{C.~S. Fischer}, \bibinfo{title}{{Baryons as relativistic
  three-quark bound states}}, \bibinfo{journal}{Prog. Part. Nucl. Phys.}
  \bibinfo{volume}{91} (\bibinfo{year}{2016}) \bibinfo{pages}{1--100}.

\bibitem[{Qin and Roberts(2020)}]{Qin:2020rad}
\bibinfo{author}{S.-X. Qin}, \bibinfo{author}{C.~D. Roberts},
  \bibinfo{title}{{Impressions of the Continuum Bound State Problem in QCD}},
  \bibinfo{journal}{Chin. Phys. Lett.}
  \bibinfo{volume}{37}~(\bibinfo{number}{12}) (\bibinfo{year}{2020})
  \bibinfo{pages}{121201}.

\bibitem[{Roberts et~al.(2021)Roberts, Richards, Horn, and
  Chang}]{Roberts:2021nhw}
\bibinfo{author}{C.~D. Roberts}, \bibinfo{author}{D.~G. Richards},
  \bibinfo{author}{T.~Horn}, \bibinfo{author}{L.~Chang},
  \bibinfo{title}{{Insights into the emergence of mass from studies of pion and
  kaon structure}}, \bibinfo{journal}{Prog. Part. Nucl. Phys.}
  \bibinfo{volume}{120} (\bibinfo{year}{2021}) \bibinfo{pages}{103883}.

\bibitem[{Binosi(2022)}]{Binosi:2022djx}
\bibinfo{author}{D.~Binosi}, \bibinfo{title}{{Emergent Hadron Mass in Strong
  Dynamics}}, \bibinfo{journal}{Few Body Syst.}
  \bibinfo{volume}{63}~(\bibinfo{number}{2}) (\bibinfo{year}{2022})
  \bibinfo{pages}{42}.

\bibitem[{Roberts(2023)}]{Roberts:2022rxm}
\bibinfo{author}{C.~D. Roberts}, \bibinfo{title}{{Origin of the Proton Mass}},
  \bibinfo{journal}{EPJ Web Conf.} \bibinfo{volume}{282} (\bibinfo{year}{2023})
  \bibinfo{pages}{01006}.

\bibitem[{Papavassiliou(2022)}]{Papavassiliou:2022wrb}
\bibinfo{author}{J.~Papavassiliou}, \bibinfo{title}{{Emergence of mass in the
  gauge sector of QCD}}, \bibinfo{journal}{Chin. Phys. C}
  \bibinfo{volume}{46}~(\bibinfo{number}{11}) (\bibinfo{year}{2022})
  \bibinfo{pages}{112001}.

\bibitem[{Ding et~al.(2023)Ding, Roberts, and Schmidt}]{Ding:2022ows}
\bibinfo{author}{M.~Ding}, \bibinfo{author}{C.~D. Roberts},
  \bibinfo{author}{S.~M. Schmidt}, \bibinfo{title}{{Emergence of Hadron Mass
  and Structure}}, \bibinfo{journal}{Particles}
  \bibinfo{volume}{6}~(\bibinfo{number}{1}) (\bibinfo{year}{2023})
  \bibinfo{pages}{57--120}.

\bibitem[{Salm\`e(2022)}]{Salme:2022eoy}
\bibinfo{author}{G.~Salm\`e}, \bibinfo{title}{{Explaining mass and spin in the
  visible matter: the next challenge}}, \bibinfo{journal}{J. Phys. Conf. Ser.}
  \bibinfo{volume}{2340}~(\bibinfo{number}{1}) (\bibinfo{year}{2022})
  \bibinfo{pages}{012011}.

\bibitem[{Ferreira and Papavassiliou(2023)}]{Ferreira:2023fva}
\bibinfo{author}{M.~N. Ferreira}, \bibinfo{author}{J.~Papavassiliou},
  \bibinfo{title}{{Gauge Sector Dynamics in QCD}}, \bibinfo{journal}{Particles}
  \bibinfo{volume}{6}~(\bibinfo{number}{1}) (\bibinfo{year}{2023})
  \bibinfo{pages}{312--363}.

\bibitem[{Carman et~al.(2023)Carman, Gothe, Mokeev, and
  Roberts}]{Carman:2023zke}
\bibinfo{author}{D.~S. Carman}, \bibinfo{author}{R.~W. Gothe},
  \bibinfo{author}{V.~I. Mokeev}, \bibinfo{author}{C.~D. Roberts},
  \bibinfo{title}{{Nucleon Resonance Electroexcitation Amplitudes and Emergent
  Hadron Mass}}, \bibinfo{journal}{Particles}
  \bibinfo{volume}{6}~(\bibinfo{number}{1}) (\bibinfo{year}{2023})
  \bibinfo{pages}{416--439}.

\bibitem[{Ding et~al.(2020)Ding, Raya, Binosi, Chang, Roberts, and
  Schmidt}]{Ding:2019qlr}
\bibinfo{author}{M.~Ding}, \bibinfo{author}{K.~Raya},
  \bibinfo{author}{D.~Binosi}, \bibinfo{author}{L.~Chang},
  \bibinfo{author}{C.~D. Roberts}, \bibinfo{author}{S.~M. Schmidt},
  \bibinfo{title}{{Drawing insights from pion parton distributions}},
  \bibinfo{journal}{Chin. Phys. C (Lett.)} \bibinfo{volume}{44}
  (\bibinfo{year}{2020}) \bibinfo{pages}{031002}.

\bibitem[{Cui et~al.(2021)Cui, Ding, Gao, Raya, Binosi, Chang, Roberts,
  Rodr\'{\i}guez-Quintero, and Schmidt}]{Cui:2020dlm}
\bibinfo{author}{Z.-F. Cui}, \bibinfo{author}{M.~Ding},
  \bibinfo{author}{F.~Gao}, \bibinfo{author}{K.~Raya},
  \bibinfo{author}{D.~Binosi}, \bibinfo{author}{L.~Chang},
  \bibinfo{author}{C.~D. Roberts},
  \bibinfo{author}{J.~Rodr\'{\i}guez-Quintero}, \bibinfo{author}{S.~M.
  Schmidt}, \bibinfo{title}{{Higgs modulation of emergent mass as revealed in
  kaon and pion parton distributions}}, \bibinfo{journal}{Eur. Phys. J. A
  (Lett.)} \bibinfo{volume}{57}~(\bibinfo{number}{1}) (\bibinfo{year}{2021})
  \bibinfo{pages}{5}.

\bibitem[{Cui et~al.(2020{\natexlab{a}})Cui, Ding, Gao, Raya, Binosi, Chang,
  Roberts, Rodr\'{\i}guez-Quintero, and Schmidt}]{Cui:2020tdf}
\bibinfo{author}{Z.-F. Cui}, \bibinfo{author}{M.~Ding},
  \bibinfo{author}{F.~Gao}, \bibinfo{author}{K.~Raya},
  \bibinfo{author}{D.~Binosi}, \bibinfo{author}{L.~Chang},
  \bibinfo{author}{C.~D. Roberts},
  \bibinfo{author}{J.~Rodr\'{\i}guez-Quintero}, \bibinfo{author}{S.~M.
  Schmidt}, \bibinfo{title}{{Kaon and pion parton distributions}},
  \bibinfo{journal}{Eur. Phys. J. C} \bibinfo{volume}{80}
  (\bibinfo{year}{2020}{\natexlab{a}}) \bibinfo{pages}{1064}.

\bibitem[{Roberts(2017)}]{Roberts:2016vyn}
\bibinfo{author}{C.~D. Roberts}, \bibinfo{title}{{Perspective on the origin of
  hadron masses}}, \bibinfo{journal}{Few Body Syst.} \bibinfo{volume}{58}
  (\bibinfo{year}{2017}) \bibinfo{pages}{5}.

\bibitem[{Cahill et~al.(1987)Cahill, Roberts, and Praschifka}]{Cahill:1987qr}
\bibinfo{author}{R.~T. Cahill}, \bibinfo{author}{C.~D. Roberts},
  \bibinfo{author}{J.~Praschifka}, \bibinfo{title}{{Calculation of diquark
  masses in QCD}}, \bibinfo{journal}{Phys. Rev. D} \bibinfo{volume}{36}
  (\bibinfo{year}{1987}) \bibinfo{pages}{2804}.

\bibitem[{Barabanov et~al.(2021)}]{Barabanov:2020jvn}
\bibinfo{author}{M.~Y. Barabanov}, et~al., \bibinfo{title}{{Diquark
  Correlations in Hadron Physics: Origin, Impact and Evidence}},
  \bibinfo{journal}{Prog. Part. Nucl. Phys.} \bibinfo{volume}{116}
  (\bibinfo{year}{2021}) \bibinfo{pages}{103835}.

\bibitem[{Cahill et~al.(1989)Cahill, Roberts, and Praschifka}]{Cahill:1988dx}
\bibinfo{author}{R.~T. Cahill}, \bibinfo{author}{C.~D. Roberts},
  \bibinfo{author}{J.~Praschifka}, \bibinfo{title}{{Baryon structure and QCD}},
  \bibinfo{journal}{Austral. J. Phys.} \bibinfo{volume}{42}
  (\bibinfo{year}{1989}) \bibinfo{pages}{129--145}.

\bibitem[{Reinhardt(1990)}]{Reinhardt:1989rw}
\bibinfo{author}{H.~Reinhardt}, \bibinfo{title}{{Hadronization of Quark Flavor
  Dynamics}}, \bibinfo{journal}{Phys. Lett. B} \bibinfo{volume}{244}
  (\bibinfo{year}{1990}) \bibinfo{pages}{316--326}.

\bibitem[{Efimov et~al.(1990)Efimov, Ivanov, and Lyubovitskij}]{Efimov:1990uz}
\bibinfo{author}{G.~V. Efimov}, \bibinfo{author}{M.~A. Ivanov},
  \bibinfo{author}{V.~E. Lyubovitskij}, \bibinfo{title}{{Quark - diquark
  approximation of the three quark structure of baryons in the quark
  confinement model}}, \bibinfo{journal}{Z. Phys. C} \bibinfo{volume}{47}
  (\bibinfo{year}{1990}) \bibinfo{pages}{583--594}.

\bibitem[{Mezrag et~al.(2018)Mezrag, Segovia, Chang, and
  Roberts}]{Mezrag:2017znp}
\bibinfo{author}{C.~Mezrag}, \bibinfo{author}{J.~Segovia},
  \bibinfo{author}{L.~Chang}, \bibinfo{author}{C.~D. Roberts},
  \bibinfo{title}{{Parton distribution amplitudes: Revealing correlations
  within the proton and Roper}}, \bibinfo{journal}{Phys. Lett. B}
  \bibinfo{volume}{783} (\bibinfo{year}{2018}) \bibinfo{pages}{263--267}.

\bibitem[{Liu et~al.(2022)Liu, Chen, Lu, Roberts, and Segovia}]{Liu:2022ndb}
\bibinfo{author}{L.~Liu}, \bibinfo{author}{C.~Chen}, \bibinfo{author}{Y.~Lu},
  \bibinfo{author}{C.~D. Roberts}, \bibinfo{author}{J.~Segovia},
  \bibinfo{title}{{Composition of low-lying $J=\tfrac{3}{2}^\pm$
  \ensuremath{\Delta}-baryons}}, \bibinfo{journal}{Phys. Rev. D}
  \bibinfo{volume}{105}~(\bibinfo{number}{11}) (\bibinfo{year}{2022})
  \bibinfo{pages}{114047}.

\bibitem[{Chang et~al.(2022)Chang, Gao, and Roberts}]{Chang:2022jri}
\bibinfo{author}{L.~Chang}, \bibinfo{author}{F.~Gao}, \bibinfo{author}{C.~D.
  Roberts}, \bibinfo{title}{{Parton distributions of light quarks and
  antiquarks in the proton}}, \bibinfo{journal}{Phys. Lett. B}
  \bibinfo{volume}{829} (\bibinfo{year}{2022}) \bibinfo{pages}{137078}.

\bibitem[{Workman et~al.(2022)}]{Workman:2022ynf}
\bibinfo{author}{R.~L. Workman}, et~al., \bibinfo{title}{{Review of Particle
  Physics}}, \bibinfo{journal}{PTEP} \bibinfo{volume}{2022}
  (\bibinfo{year}{2022}) \bibinfo{pages}{083C01}.

\bibitem[{Chen and Roberts(2022)}]{ChenChen:2022qpy}
\bibinfo{author}{C.~Chen}, \bibinfo{author}{C.~D. Roberts},
  \bibinfo{title}{{Nucleon axial form factor at large momentum transfers}},
  \bibinfo{journal}{Eur. Phys. J. A} \bibinfo{volume}{58}
  (\bibinfo{year}{2022}) \bibinfo{pages}{206}.

\bibitem[{Cheng et~al.(2022)Cheng, Serna, Yao, Chen, Cui, and
  Roberts}]{Cheng:2022jxe}
\bibinfo{author}{P.~Cheng}, \bibinfo{author}{F.~E. Serna},
  \bibinfo{author}{Z.-Q. Yao}, \bibinfo{author}{C.~Chen},
  \bibinfo{author}{Z.-F. Cui}, \bibinfo{author}{C.~D. Roberts},
  \bibinfo{title}{{Contact interaction analysis of octet baryon axial-vector
  and pseudoscalar form factors}}, \bibinfo{journal}{Phys. Rev. D}
  \bibinfo{volume}{106}~(\bibinfo{number}{5}) (\bibinfo{year}{2022})
  \bibinfo{pages}{054031}.

\bibitem[{Chen et~al.(2021)Chen, Fischer, Roberts, and Segovia}]{Chen:2020wuq}
\bibinfo{author}{C.~Chen}, \bibinfo{author}{C.~S. Fischer},
  \bibinfo{author}{C.~D. Roberts}, \bibinfo{author}{J.~Segovia},
  \bibinfo{title}{{Form Factors of the Nucleon Axial Current}},
  \bibinfo{journal}{Phys. Lett. B} \bibinfo{volume}{815} (\bibinfo{year}{2021})
  \bibinfo{pages}{136150}.

\bibitem[{Chen et~al.(2022)Chen, Fischer, Roberts, and Segovia}]{Chen:2021guo}
\bibinfo{author}{C.~Chen}, \bibinfo{author}{C.~S. Fischer},
  \bibinfo{author}{C.~D. Roberts}, \bibinfo{author}{J.~Segovia},
  \bibinfo{title}{{Nucleon axial-vector and pseudoscalar form factors and PCAC
  relations}}, \bibinfo{journal}{Phys. Rev. D}
  \bibinfo{volume}{105}~(\bibinfo{number}{9}) (\bibinfo{year}{2022})
  \bibinfo{pages}{094022}.

\bibitem[{Liu et~al.(2020)Liu, Sufian, de~T\'eramond, Dosch, Brodsky, and
  Deur}]{Liu:2019vsn}
\bibinfo{author}{T.~Liu}, \bibinfo{author}{R.~S. Sufian},
  \bibinfo{author}{G.~F. de~T\'eramond}, \bibinfo{author}{H.~G. Dosch},
  \bibinfo{author}{S.~J. Brodsky}, \bibinfo{author}{A.~Deur},
  \bibinfo{title}{{Unified Description of Polarized and Unpolarized Quark
  Distributions in the Proton}}, \bibinfo{journal}{Phys. Rev. Lett.}
  \bibinfo{volume}{124}~(\bibinfo{number}{8}) (\bibinfo{year}{2020})
  \bibinfo{pages}{082003}.

\bibitem[{Han et~al.(2022)Han, Xie, Wang, and Chen}]{Han:2021dkc}
\bibinfo{author}{C.~Han}, \bibinfo{author}{G.~Xie}, \bibinfo{author}{R.~Wang},
  \bibinfo{author}{X.~Chen}, \bibinfo{title}{{An analysis of polarized parton
  distribution functions with nonlinear QCD evolution equations}},
  \bibinfo{journal}{Nucl. Phys. B} \bibinfo{volume}{985} (\bibinfo{year}{2022})
  \bibinfo{pages}{116012}.

\bibitem[{Brodsky et~al.(1995)Brodsky, Burkardt, and Schmidt}]{Brodsky:1994kg}
\bibinfo{author}{S.~J. Brodsky}, \bibinfo{author}{M.~Burkardt},
  \bibinfo{author}{I.~Schmidt}, \bibinfo{title}{{Perturbative QCD constraints
  on the shape of polarized quark and gluon distributions}},
  \bibinfo{journal}{Nucl. Phys. B} \bibinfo{volume}{441} (\bibinfo{year}{1995})
  \bibinfo{pages}{197--214}.

\bibitem[{Brisudova et~al.(2000)Brisudova, Burakovsky, and
  Goldman}]{Brisudova:1999ut}
\bibinfo{author}{M.~M. Brisudova}, \bibinfo{author}{L.~Burakovsky},
  \bibinfo{author}{J.~T. Goldman}, \bibinfo{title}{{Effective functional form
  of Regge trajectories}}, \bibinfo{journal}{Phys. Rev. D} \bibinfo{volume}{61}
  (\bibinfo{year}{2000}) \bibinfo{pages}{054013}.

\bibitem[{Roberts et~al.(2013)Roberts, Holt, and Schmidt}]{Roberts:2013mja}
\bibinfo{author}{C.~D. Roberts}, \bibinfo{author}{R.~J. Holt},
  \bibinfo{author}{S.~M. Schmidt}, \bibinfo{title}{{Nucleon spin structure at
  very high $x$}}, \bibinfo{journal}{Phys. Lett. B} \bibinfo{volume}{727}
  (\bibinfo{year}{2013}) \bibinfo{pages}{249--254}.

\bibitem[{Close and Thomas(1988)}]{Close:1988br}
\bibinfo{author}{F.~E. Close}, \bibinfo{author}{A.~W. Thomas},
  \bibinfo{title}{{The Spin and Flavor Dependence of Parton Distribution
  Functions}}, \bibinfo{journal}{Phys. Lett. B} \bibinfo{volume}{212}
  (\bibinfo{year}{1988}) \bibinfo{pages}{227}.

\bibitem[{Hughes and Voss(1999)}]{Hughes:1999wr}
\bibinfo{author}{E.~W. Hughes}, \bibinfo{author}{R.~Voss},
  \bibinfo{title}{{Spin structure functions}}, \bibinfo{journal}{Ann. Rev.
  Nucl. Part. Sci.} \bibinfo{volume}{49} (\bibinfo{year}{1999})
  \bibinfo{pages}{303--339}.

\bibitem[{Farrar and Jackson(1975)}]{Farrar:1975yb}
\bibinfo{author}{G.~R. Farrar}, \bibinfo{author}{D.~R. Jackson},
  \bibinfo{title}{{Pion and Nucleon Structure Functions Near $x=1$}},
  \bibinfo{journal}{Phys. Rev. Lett.} \bibinfo{volume}{35}
  (\bibinfo{year}{1975}) \bibinfo{pages}{1416}.

\bibitem[{Holt and Roberts(2010)}]{Holt:2010vj}
\bibinfo{author}{R.~J. Holt}, \bibinfo{author}{C.~D. Roberts},
  \bibinfo{title}{{Distribution Functions of the Nucleon and Pion in the
  Valence Region}}, \bibinfo{journal}{Rev. Mod. Phys.} \bibinfo{volume}{82}
  (\bibinfo{year}{2010}) \bibinfo{pages}{2991--3044}.

\bibitem[{Dokshitzer(1977)}]{Dokshitzer:1977sg}
\bibinfo{author}{Y.~L. Dokshitzer}, \bibinfo{title}{Calculation of the
  Structure Functions for Deep Inelastic Scattering and $e^+$ $e^-$
  Annihilation by Perturbation Theory in Quantum Chromodynamics. ({\mbox {I}n
  {R}ussian})}, \bibinfo{journal}{Sov. Phys. JETP} \bibinfo{volume}{46}
  (\bibinfo{year}{1977}) \bibinfo{pages}{641--653}.

\bibitem[{Gribov and Lipatov(1971)}]{Gribov:1971zn}
\bibinfo{author}{V.~N. Gribov}, \bibinfo{author}{L.~N. Lipatov},
  \bibinfo{title}{{Deep inelastic electron scattering in perturbation theory}},
  \bibinfo{journal}{Phys. Lett. B} \bibinfo{volume}{37} (\bibinfo{year}{1971})
  \bibinfo{pages}{78--80}.

\bibitem[{Lipatov(1975)}]{Lipatov:1974qm}
\bibinfo{author}{L.~N. Lipatov}, \bibinfo{title}{{The parton model and
  perturbation theory}}, \bibinfo{journal}{Sov. J. Nucl. Phys.}
  \bibinfo{volume}{20} (\bibinfo{year}{1975}) \bibinfo{pages}{94--102}.

\bibitem[{Altarelli and Parisi(1977)}]{Altarelli:1977zs}
\bibinfo{author}{G.~Altarelli}, \bibinfo{author}{G.~Parisi},
  \bibinfo{title}{{Asymptotic Freedom in Parton Language}},
  \bibinfo{journal}{Nucl. Phys. B} \bibinfo{volume}{126} (\bibinfo{year}{1977})
  \bibinfo{pages}{298--318}.

\bibitem[{Airapetian et~al.(2005)}]{HERMES:2004zsh}
\bibinfo{author}{A.~Airapetian}, et~al., \bibinfo{title}{{Quark helicity
  distributions in the nucleon for up, down, and strange quarks from
  semi-inclusive deep-inelastic scattering}}, \bibinfo{journal}{Phys. Rev. D}
  \bibinfo{volume}{71} (\bibinfo{year}{2005}) \bibinfo{pages}{012003}.

\bibitem[{Alekseev et~al.(2010)}]{COMPASS:2010hwr}
\bibinfo{author}{M.~G. Alekseev}, et~al., \bibinfo{title}{{Quark helicity
  distributions from longitudinal spin asymmetries in muon-proton and
  muon-deuteron scattering}}, \bibinfo{journal}{Phys. Lett. B}
  \bibinfo{volume}{693} (\bibinfo{year}{2010}) \bibinfo{pages}{227--235}.

\bibitem[{Dharmawardane et~al.(2006)}]{CLAS:2006ozz}
\bibinfo{author}{K.~V. Dharmawardane}, et~al., \bibinfo{title}{{Measurement of
  the $x$- and $Q^2$-dependence of the asymmetry $A_1$ on the nucleon}},
  \bibinfo{journal}{Phys. Lett. B} \bibinfo{volume}{641} (\bibinfo{year}{2006})
  \bibinfo{pages}{11--17}.

\bibitem[{Prok et~al.(2009)}]{CLAS:2008xos}
\bibinfo{author}{Y.~Prok}, et~al., \bibinfo{title}{{Moments of the Spin
  Structure Functions $g_1^p$ and $g_1^d$ for $0.05 < Q^2 < 3$-GeV$^2$}},
  \bibinfo{journal}{Phys. Lett. B} \bibinfo{volume}{672} (\bibinfo{year}{2009})
  \bibinfo{pages}{12--16}.

\bibitem[{Guler et~al.(2015)}]{CLAS:2015otq}
\bibinfo{author}{N.~Guler}, et~al., \bibinfo{title}{{Precise determination of
  the deuteron spin structure at low to moderate $Q^2$ with CLAS and extraction
  of the neutron contribution}}, \bibinfo{journal}{Phys. Rev. C}
  \bibinfo{volume}{92}~(\bibinfo{number}{5}) (\bibinfo{year}{2015})
  \bibinfo{pages}{055201}.

\bibitem[{Fersch et~al.(2017)}]{CLAS:2017qga}
\bibinfo{author}{R.~Fersch}, et~al., \bibinfo{title}{{Determination of the
  Proton Spin Structure Functions for $0.05 < Q^{2} < 5\, {\rm GeV}^{2}$ using
  CLAS}}, \bibinfo{journal}{Phys. Rev. C}
  \bibinfo{volume}{96}~(\bibinfo{number}{6}) (\bibinfo{year}{2017})
  \bibinfo{pages}{065208}.

\bibitem[{Parno et~al.(2015)}]{JeffersonLabHallA:2014mam}
\bibinfo{author}{D.~S. Parno}, et~al., \bibinfo{title}{{Precision Measurements
  of $A_1^n$ in the Deep Inelastic Regime}}, \bibinfo{journal}{Phys. Lett. B}
  \bibinfo{volume}{744} (\bibinfo{year}{2015}) \bibinfo{pages}{309--314}.

\bibitem[{Zheng et~al.(2004{\natexlab{a}})}]{JeffersonLabHallA:2003joy}
\bibinfo{author}{X.~Zheng}, et~al., \bibinfo{title}{{Precision measurement of
  the neutron spin asymmetry $A_1^N$ and spin flavor decomposition in the
  valence quark region}}, \bibinfo{journal}{Phys. Rev. Lett.}
  \bibinfo{volume}{92} (\bibinfo{year}{2004}{\natexlab{a}})
  \bibinfo{pages}{012004}.

\bibitem[{Zheng et~al.(2004{\natexlab{b}})}]{JeffersonLabHallA:2004tea}
\bibinfo{author}{X.~Zheng}, et~al., \bibinfo{title}{{Precision measurement of
  the neutron spin asymmetries and spin-dependent structure functions in the
  valence quark region}}, \bibinfo{journal}{Phys. Rev. C} \bibinfo{volume}{70}
  (\bibinfo{year}{2004}{\natexlab{b}}) \bibinfo{pages}{065207}.

\bibitem[{Deur et~al.(2019)Deur, Brodsky, and De~T\'eramond}]{Deur:2018roz}
\bibinfo{author}{A.~Deur}, \bibinfo{author}{S.~J. Brodsky},
  \bibinfo{author}{G.~F. De~T\'eramond}, \bibinfo{title}{{The Spin Structure of
  the Nucleon}}, \bibinfo{journal}{Rept. Prog. Phys.}
  \bibinfo{volume}{82}~(\bibinfo{number}{076201}).

\bibitem[{Xu et~al.(2021)Xu, Mondal, Lan, Zhao, Li, and Vary}]{Xu:2021wwj}
\bibinfo{author}{S.~Xu}, \bibinfo{author}{C.~Mondal}, \bibinfo{author}{J.~Lan},
  \bibinfo{author}{X.~Zhao}, \bibinfo{author}{Y.~Li}, \bibinfo{author}{J.~P.
  Vary}, \bibinfo{title}{{Nucleon structure from basis light-front
  quantization}}, \bibinfo{journal}{Phys. Rev. D}
  \bibinfo{volume}{104}~(\bibinfo{number}{9}) (\bibinfo{year}{2021})
  \bibinfo{pages}{094036}.

\bibitem[{Xu et~al.(2022)Xu, Mondal, Zhao, Li, and Vary}]{Xu:2022abw}
\bibinfo{author}{S.~Xu}, \bibinfo{author}{C.~Mondal},
  \bibinfo{author}{X.~Zhao}, \bibinfo{author}{Y.~Li}, \bibinfo{author}{J.~P.
  Vary}, \bibinfo{title}{{Nucleon spin decomposition with one dynamical gluon
  -- arXiv:2209.08584 [hep-ph]}} .

\bibitem[{Ethier and Nocera(2020)}]{Ethier:2020way}
\bibinfo{author}{J.~J. Ethier}, \bibinfo{author}{E.~R. Nocera},
  \bibinfo{title}{{Parton Distributions in Nucleons and Nuclei}},
  \bibinfo{journal}{Ann. Rev. Nucl. Part. Sci.} \bibinfo{volume}{70}
  (\bibinfo{year}{2020}) \bibinfo{pages}{43--76}.

\bibitem[{Kuhn et~al.( 109)}]{E1206109}
\bibinfo{author}{S.~Kuhn}, et~al., \bibinfo{note}{{\emph{The Longitudinal Spin
  Structure of the Nucleon}}}, \bibinfo{year}{CLAS Collaboration (E12-06-109)}.

\bibitem[{Zheng et~al.(tion)}]{Zheng:2006}
\bibinfo{author}{X.~Zheng}, et~al., \bibinfo{note}{{\emph{Measurement of
  Neutron Spin Asymmetry A$_1^n$ in the Valence Quark Region using an 11 GeV
  Beam and a Polarized $^3$He Target in Hall C}}}, \bibinfo{year}{E12-06-110
  Collaboration}.

\bibitem[{Bali et~al.(2016)}]{Bali:2015ykx}
\bibinfo{author}{G.~S. Bali}, et~al., \bibinfo{title}{{Light-cone distribution
  amplitudes of the baryon octet}}, \bibinfo{journal}{JHEP}
  \bibinfo{volume}{02} (\bibinfo{year}{2016}) \bibinfo{pages}{070}.

\bibitem[{Lepage and Brodsky(1980)}]{Lepage:1980fj}
\bibinfo{author}{G.~P. Lepage}, \bibinfo{author}{S.~J. Brodsky},
  \bibinfo{title}{{Exclusive Processes in Perturbative Quantum
  Chromodynamics}}, \bibinfo{journal}{Phys. Rev. D} \bibinfo{volume}{22}
  (\bibinfo{year}{1980}) \bibinfo{pages}{2157--2198}.

\bibitem[{Lu et~al.(2022)Lu, Chang, Raya, Roberts, and
  Rodr\'\i{}guez-Quintero}]{Lu:2022cjx}
\bibinfo{author}{Y.~Lu}, \bibinfo{author}{L.~Chang}, \bibinfo{author}{K.~Raya},
  \bibinfo{author}{C.~D. Roberts},
  \bibinfo{author}{J.~Rodr\'\i{}guez-Quintero}, \bibinfo{title}{{Proton and
  pion distribution functions in counterpoint}}, \bibinfo{journal}{Phys. Lett.
  B} \bibinfo{volume}{830} (\bibinfo{year}{2022}) \bibinfo{pages}{137130}.

\bibitem[{Abrams et~al.(2022)}]{Abrams:2021xum}
\bibinfo{author}{D.~Abrams}, et~al., \bibinfo{title}{{Measurement of the
  Nucleon $F^n_2/F^p_2$ Structure Function Ratio by the Jefferson Lab MARATHON
  Tritium/Helium-3 Deep Inelastic Scattering Experiment}},
  \bibinfo{journal}{Phys. Rev. Lett.}
  \bibinfo{volume}{128}~(\bibinfo{number}{13}) (\bibinfo{year}{2022})
  \bibinfo{pages}{132003}.

\bibitem[{Cui et~al.(2022{\natexlab{a}})Cui, Gao, Binosi, Chang, Roberts, and
  Schmidt}]{Cui:2021gzg}
\bibinfo{author}{Z.-F. Cui}, \bibinfo{author}{F.~Gao},
  \bibinfo{author}{D.~Binosi}, \bibinfo{author}{L.~Chang},
  \bibinfo{author}{C.~D. Roberts}, \bibinfo{author}{S.~M. Schmidt},
  \bibinfo{title}{{Valence quark ratio in the proton}}, \bibinfo{journal}{Chin.
  Phys. Lett. \emph{Express}} \bibinfo{volume}{39}~(\bibinfo{number}{04})
  (\bibinfo{year}{2022}{\natexlab{a}}) \bibinfo{pages}{041401}.

\bibitem[{Cui et~al.(2020{\natexlab{b}})Cui, Zhang, Binosi, de~Soto, Mezrag,
  Papavassiliou, Roberts, Rodr{\'{\i}}guez-Quintero, Segovia, and
  Zafeiropoulos}]{Cui:2019dwv}
\bibinfo{author}{Z.-F. Cui}, \bibinfo{author}{J.-L. Zhang},
  \bibinfo{author}{D.~Binosi}, \bibinfo{author}{F.~de~Soto},
  \bibinfo{author}{C.~Mezrag}, \bibinfo{author}{J.~Papavassiliou},
  \bibinfo{author}{C.~D. Roberts},
  \bibinfo{author}{J.~Rodr{\'{\i}}guez-Quintero}, \bibinfo{author}{J.~Segovia},
  \bibinfo{author}{S.~Zafeiropoulos}, \bibinfo{title}{{Effective charge from
  lattice QCD}}, \bibinfo{journal}{Chin. Phys. C} \bibinfo{volume}{44}
  (\bibinfo{year}{2020}{\natexlab{b}}) \bibinfo{pages}{083102}.

\bibitem[{Raya et~al.(2022)Raya, Cui, Chang, Morgado, Roberts, and
  Rodr{\'{\i}}guez-Quintero}]{Raya:2021zrz}
\bibinfo{author}{K.~Raya}, \bibinfo{author}{Z.-F. Cui},
  \bibinfo{author}{L.~Chang}, \bibinfo{author}{J.-M. Morgado},
  \bibinfo{author}{C.~D. Roberts},
  \bibinfo{author}{J.~Rodr{\'{\i}}guez-Quintero}, \bibinfo{title}{{Revealing
  pion and kaon structure via generalised parton distributions}},
  \bibinfo{journal}{Chin. Phys. C} \bibinfo{volume}{46}~(\bibinfo{number}{26})
  (\bibinfo{year}{2022}) \bibinfo{pages}{013105}.

\bibitem[{Cui et~al.(2022{\natexlab{b}})Cui, Ding, Morgado, Raya, Binosi,
  Chang, Papavassiliou, Roberts, Rodr\'\i{}guez-Quintero, and
  Schmidt}]{Cui:2021mom}
\bibinfo{author}{Z.~F. Cui}, \bibinfo{author}{M.~Ding}, \bibinfo{author}{J.~M.
  Morgado}, \bibinfo{author}{K.~Raya}, \bibinfo{author}{D.~Binosi},
  \bibinfo{author}{L.~Chang}, \bibinfo{author}{J.~Papavassiliou},
  \bibinfo{author}{C.~D. Roberts},
  \bibinfo{author}{J.~Rodr\'\i{}guez-Quintero}, \bibinfo{author}{S.~M.
  Schmidt}, \bibinfo{title}{{Concerning pion parton distributions}},
  \bibinfo{journal}{Eur. Phys. J. A} \bibinfo{volume}{58}~(\bibinfo{number}{1})
  (\bibinfo{year}{2022}{\natexlab{b}}) \bibinfo{pages}{10}.

\bibitem[{Cui et~al.(2022{\natexlab{c}})Cui, Ding, Morgado, Raya, Binosi,
  Chang, De~Soto, Roberts, Rodr\'\i{}guez-Quintero, and Schmidt}]{Cui:2022bxn}
\bibinfo{author}{Z.~F. Cui}, \bibinfo{author}{M.~Ding}, \bibinfo{author}{J.~M.
  Morgado}, \bibinfo{author}{K.~Raya}, \bibinfo{author}{D.~Binosi},
  \bibinfo{author}{L.~Chang}, \bibinfo{author}{F.~De~Soto},
  \bibinfo{author}{C.~D. Roberts},
  \bibinfo{author}{J.~Rodr\'\i{}guez-Quintero}, \bibinfo{author}{S.~M.
  Schmidt}, \bibinfo{title}{{Emergence of pion parton distributions}},
  \bibinfo{journal}{Phys. Rev. D} \bibinfo{volume}{105}~(\bibinfo{number}{9})
  (\bibinfo{year}{2022}{\natexlab{c}}) \bibinfo{pages}{L091502}.

\bibitem[{Xu et~al.(2023)Xu, Raya, Cui, Roberts, and
  Rodr\'\i{}guez-Quintero}]{Xu:2023bwv}
\bibinfo{author}{Y.-Z. Xu}, \bibinfo{author}{K.~Raya}, \bibinfo{author}{Z.-F.
  Cui}, \bibinfo{author}{C.~D. Roberts},
  \bibinfo{author}{J.~Rodr\'\i{}guez-Quintero}, \bibinfo{title}{{Empirical
  Determination of the Pion Mass Distribution}}, \bibinfo{journal}{Chin. Phys.
  Lett. \emph{Express}} \bibinfo{volume}{40}~(\bibinfo{number}{4})
  (\bibinfo{year}{2023}) \bibinfo{pages}{041201}.

\bibitem[{Grunberg(1980)}]{Grunberg:1980ja}
\bibinfo{author}{G.~Grunberg}, \bibinfo{title}{{Renormalization Group Improved
  Perturbative QCD}}, \bibinfo{journal}{Phys. Lett. B} \bibinfo{volume}{95}
  (\bibinfo{year}{1980}) \bibinfo{pages}{70}, \bibinfo{note}{[Erratum: Phys.
  Lett. B 110, 501 (1982)]}.

\bibitem[{Grunberg(1984)}]{Grunberg:1982fw}
\bibinfo{author}{G.~Grunberg}, \bibinfo{title}{{Renormalization Scheme
  Independent QCD and QED: The Method of Effective Charges}},
  \bibinfo{journal}{Phys. Rev. D} \bibinfo{volume}{29} (\bibinfo{year}{1984})
  \bibinfo{pages}{2315}.

\bibitem[{Deur et~al.(2016)Deur, Brodsky, and de~Teramond}]{Deur:2016tte}
\bibinfo{author}{A.~Deur}, \bibinfo{author}{S.~J. Brodsky},
  \bibinfo{author}{G.~F. de~Teramond}, \bibinfo{title}{{The QCD Running
  Coupling}}, \bibinfo{journal}{Prog. Part. Nucl. Phys.} \bibinfo{volume}{90}
  (\bibinfo{year}{2016}) \bibinfo{pages}{1--74}.

\bibitem[{Deur et~al.(2022)Deur, Burkert, Chen, and Korsch}]{Deur:2022msf}
\bibinfo{author}{A.~Deur}, \bibinfo{author}{V.~Burkert}, \bibinfo{author}{J.~P.
  Chen}, \bibinfo{author}{W.~Korsch}, \bibinfo{title}{{Experimental
  determination of the QCD effective charge $\alpha_{g_1}(Q)$}},
  \bibinfo{journal}{Particles} \bibinfo{volume}{5}~(\bibinfo{number}{2})
  (\bibinfo{year}{2022}) \bibinfo{pages}{171--179}.

\bibitem[{Deur et~al.(2023)Deur, Brodsky, and Roberts}]{Deur:2023dzc}
\bibinfo{author}{A.~Deur}, \bibinfo{author}{S.~J. Brodsky},
  \bibinfo{author}{C.~D. Roberts}, \bibinfo{title}{{QCD Running Couplings and
  Effective Charges -- arXiv:2303.00723 [hep-ph]}} .

\bibitem[{Dove et~al.(2021)}]{SeaQuest:2021zxb}
\bibinfo{author}{J.~Dove}, et~al., \bibinfo{title}{{The asymmetry of antimatter
  in the proton}}, \bibinfo{journal}{Nature}
  \bibinfo{volume}{590}~(\bibinfo{number}{7847}) (\bibinfo{year}{2021})
  \bibinfo{pages}{561--565}.

\bibitem[{Prok et~al.(2014)}]{CLAS:2014qtg}
\bibinfo{author}{Y.~Prok}, et~al., \bibinfo{title}{{Precision measurements of
  $g_1$ of the proton and the deuteron with 6 GeV electrons}},
  \bibinfo{journal}{Phys. Rev. C} \bibinfo{volume}{90}~(\bibinfo{number}{2})
  (\bibinfo{year}{2014}) \bibinfo{pages}{025212}.

\bibitem[{Anthony et~al.(1999)}]{E155:1999pwm}
\bibinfo{author}{P.~L. Anthony}, et~al., \bibinfo{title}{{Measurement of the
  deuteron spin structure function $g_1^d(x)$ for 1-(GeV/c)$^2< Q^2<
  40$-(GeV/c)$^2$}}, \bibinfo{journal}{Phys. Lett. B} \bibinfo{volume}{463}
  (\bibinfo{year}{1999}) \bibinfo{pages}{339--345}.

\bibitem[{Anthony et~al.(2000)}]{E155:2000qdr}
\bibinfo{author}{P.~L. Anthony}, et~al., \bibinfo{title}{{Measurements of the
  $Q^2$ dependence of the proton and neutron spin structure functions $g_1^p$
  and $g_1^n$}}, \bibinfo{journal}{Phys. Lett. B} \bibinfo{volume}{493}
  (\bibinfo{year}{2000}) \bibinfo{pages}{19--28}.

\bibitem[{Airapetian et~al.(1998)}]{HERMES:1998cbu}
\bibinfo{author}{A.~Airapetian}, et~al., \bibinfo{title}{{Measurement of the
  proton spin structure function g1(p) with a pure hydrogen target}},
  \bibinfo{journal}{Phys. Lett. B} \bibinfo{volume}{442} (\bibinfo{year}{1998})
  \bibinfo{pages}{484--492}.

\bibitem[{Abe et~al.(1995)}]{E143:1995clm}
\bibinfo{author}{K.~Abe}, et~al., \bibinfo{title}{{Measurements of the $Q^2$
  dependence of the proton and deuteron spin structure functions $g_1^p$ and
  $g_1^d$}}, \bibinfo{journal}{Phys. Lett. B} \bibinfo{volume}{364}
  (\bibinfo{year}{1995}) \bibinfo{pages}{61--68}.

\bibitem[{Abe et~al.(1997{\natexlab{a}})}]{E143:1996vck}
\bibinfo{author}{K.~Abe}, et~al., \bibinfo{title}{{Measurements of the proton
  and deuteron spin structure function $g_1$ in the resonance region}},
  \bibinfo{journal}{Phys. Rev. Lett.} \bibinfo{volume}{78}
  (\bibinfo{year}{1997}{\natexlab{a}}) \bibinfo{pages}{815--819}.

\bibitem[{Abe et~al.(1998)}]{E143:1998hbs}
\bibinfo{author}{K.~Abe}, et~al., \bibinfo{title}{{Measurements of the proton
  and deuteron spin structure functions $g_1$ and $g_2$}},
  \bibinfo{journal}{Phys. Rev. D} \bibinfo{volume}{58} (\bibinfo{year}{1998})
  \bibinfo{pages}{112003}.

\bibitem[{Adams et~al.(1994)}]{SpinMuonSMC:1994met}
\bibinfo{author}{D.~Adams}, et~al., \bibinfo{title}{{Measurement of the spin
  dependent structure function $g_1(x)$ of the proton}},
  \bibinfo{journal}{Phys. Lett. B} \bibinfo{volume}{329} (\bibinfo{year}{1994})
  \bibinfo{pages}{399--406}, \bibinfo{note}{[Erratum: Phys. Lett. B 339,
  332--333 (1994)]}.

\bibitem[{Adams et~al.(1997)}]{SpinMuonSMC:1997mkb}
\bibinfo{author}{D.~Adams}, et~al., \bibinfo{title}{{Spin structure of the
  proton from polarized inclusive deep inelastic muon - proton scattering}},
  \bibinfo{journal}{Phys. Rev. D} \bibinfo{volume}{56} (\bibinfo{year}{1997})
  \bibinfo{pages}{5330--5358}.

\bibitem[{Ackerstaff et~al.(1997)}]{HERMES:1997hjr}
\bibinfo{author}{K.~Ackerstaff}, et~al., \bibinfo{title}{{Measurement of the
  neutron spin structure function $g_1^n$ with a polarized $\,^3$He internal
  target}}, \bibinfo{journal}{Phys. Lett. B} \bibinfo{volume}{404}
  (\bibinfo{year}{1997}) \bibinfo{pages}{383--389}.

\bibitem[{Abe et~al.(1997{\natexlab{b}})}]{E154:1997xfa}
\bibinfo{author}{K.~Abe}, et~al., \bibinfo{title}{{Precision determination of
  the neutron spin structure function $g_1^n$}}, \bibinfo{journal}{Phys. Rev.
  Lett.} \bibinfo{volume}{79} (\bibinfo{year}{1997}{\natexlab{b}})
  \bibinfo{pages}{26--30}.

\bibitem[{Abe et~al.(1997{\natexlab{c}})}]{E154:1997ysl}
\bibinfo{author}{K.~Abe}, et~al., \bibinfo{title}{{Next-to-leading order QCD
  analysis of polarized deep inelastic scattering data}},
  \bibinfo{journal}{Phys. Lett. B} \bibinfo{volume}{405}
  (\bibinfo{year}{1997}{\natexlab{c}}) \bibinfo{pages}{180--190}.

\bibitem[{Anthony et~al.(1996)}]{E142:1996thl}
\bibinfo{author}{P.~L. Anthony}, et~al., \bibinfo{title}{{Deep inelastic
  scattering of polarized electrons by polarized $\,^3$He and the study of the
  neutron spin structure}}, \bibinfo{journal}{Phys. Rev. D}
  \bibinfo{volume}{54} (\bibinfo{year}{1996}) \bibinfo{pages}{6620--6650}.

\bibitem[{Ellis et~al.(1991)Ellis, Stirling, and Webber}]{Ellis:1991qj}
\bibinfo{author}{R.~K. Ellis}, \bibinfo{author}{W.~J. Stirling},
  \bibinfo{author}{B.~R. Webber}, \bibinfo{title}{{\mbox{$\;$}QCD and collider
  physics}}, \bibinfo{publisher}{Cambridge University Press, Cambridge, UK},
  \bibinfo{year}{1991}.

\bibitem[{de~Florian et~al.(2014)de~Florian, Sassot, Stratmann, and
  Vogelsang}]{deFlorian:2014yva}
\bibinfo{author}{D.~de~Florian}, \bibinfo{author}{R.~Sassot},
  \bibinfo{author}{M.~Stratmann}, \bibinfo{author}{W.~Vogelsang},
  \bibinfo{title}{{Evidence for polarization of gluons in the proton}},
  \bibinfo{journal}{Phys. Rev. Lett.}
  \bibinfo{volume}{113}~(\bibinfo{number}{1}) (\bibinfo{year}{2014})
  \bibinfo{pages}{012001}.

\bibitem[{Adolph et~al.(2017{\natexlab{a}})}]{COMPASS:2015pim}
\bibinfo{author}{C.~Adolph}, et~al., \bibinfo{title}{{Leading-order
  determination of the gluon polarisation from semi-inclusive deep inelastic
  scattering data}}, \bibinfo{journal}{Eur. Phys. J. C}
  \bibinfo{volume}{77}~(\bibinfo{number}{4})
  (\bibinfo{year}{2017}{\natexlab{a}}) \bibinfo{pages}{209}.

\bibitem[{Altarelli and Ross(1988)}]{Altarelli:1988nr}
\bibinfo{author}{G.~Altarelli}, \bibinfo{author}{G.~G. Ross},
  \bibinfo{title}{{The Anomalous Gluon Contribution to Polarized
  Leptoproduction}}, \bibinfo{journal}{Phys. Lett. B} \bibinfo{volume}{212}
  (\bibinfo{year}{1988}) \bibinfo{pages}{391--396}.

\bibitem[{Adolph et~al.(2017{\natexlab{b}})}]{COMPASS:2016jwv}
\bibinfo{author}{C.~Adolph}, et~al., \bibinfo{title}{{Final COMPASS results on
  the deuteron spin-dependent structure function $g_1^{\rm d}$ and the Bjorken
  sum rule}}, \bibinfo{journal}{Phys. Lett. B} \bibinfo{volume}{769}
  (\bibinfo{year}{2017}{\natexlab{b}}) \bibinfo{pages}{34--41}.

\bibitem[{Ji et~al.(1996)Ji, Tang, and Hoodbhoy}]{Ji:1995cu}
\bibinfo{author}{X.-D. Ji}, \bibinfo{author}{J.~Tang},
  \bibinfo{author}{P.~Hoodbhoy}, \bibinfo{title}{{The spin structure of the
  nucleon in the asymptotic limit}}, \bibinfo{journal}{Phys. Rev. Lett.}
  \bibinfo{volume}{76} (\bibinfo{year}{1996}) \bibinfo{pages}{740--743}.

\bibitem[{Chen et~al.(2011)Chen, Sun, Wang, and Goldman}]{Chen:2011gn}
\bibinfo{author}{X.-S. Chen}, \bibinfo{author}{W.-M. Sun},
  \bibinfo{author}{F.~Wang}, \bibinfo{author}{T.~Goldman},
  \bibinfo{title}{{Proper identification of the gluon spin}},
  \bibinfo{journal}{Phys. Lett. B} \bibinfo{volume}{700} (\bibinfo{year}{2011})
  \bibinfo{pages}{21--24}.

\bibitem[{Anderle et~al.(2021)}]{Anderle:2021wcy}
\bibinfo{author}{D.~P. Anderle}, et~al., \bibinfo{title}{{Electron-ion collider
  in China}}, \bibinfo{journal}{Front. Phys. (Beijing)}
  \bibinfo{volume}{16}~(\bibinfo{number}{6}) (\bibinfo{year}{2021})
  \bibinfo{pages}{64701}.

\bibitem[{Abdul~Khalek et~al.(2022)}]{AbdulKhalek:2021gbh}
\bibinfo{author}{R.~Abdul~Khalek}, et~al., \bibinfo{title}{{Science
  Requirements and Detector Concepts for the Electron-Ion Collider: EIC Yellow
  Report}}, \bibinfo{journal}{Nucl. Phys. A} \bibinfo{volume}{1026}
  (\bibinfo{year}{2022}) \bibinfo{pages}{122447}.

\end{thebibliography}

\end{document}